\documentclass[12pt]{iopart}

\usepackage{graphicx}
\usepackage{cite}
\usepackage{color}

\begin{document}

\title{Properties of Laughlin states on fractal lattices}

\author{Mani Chandra Jha$^1$ and Anne E. B. Nielsen$^2$}
\address{$^1$Max-Planck-Institut f\"ur Physik komplexer Systeme, D-01187 Dresden, Germany}
\address{$^2$Department of Physics and Astronomy, Aarhus University, DK-8000 Aarhus C, Denmark.}

\begin{abstract}
Laughlin states have recently been constructed on fractal lattices and have been shown to be topological in such systems. Some of their properties are, however, quite different from the two-dimensional case. On the Sierpinski triangle, for instance, the entanglement entropy shows oscillations as a function of particle number and does not obey the area law despite being topologically ordered, and the particle density is non-uniform in the bulk. Here, we investigate these deviant properties in greater detail on the Sierpinski triangle, and we also study the properties on the Sierpinski carpet and the T-fractal. We find that the density variations across the fractal are present for all the considered fractal lattices and for most choices of the number of particles. The size of anyons inserted into the lattice Laughlin state also varies with position on the fractal lattice. We observe that quasiholes and quasiparticles have similar sizes and that the size of the anyons typically increases with decreasing Hausdorff dimension. As opposed to periodic lattices in two dimensions, the Sierpinski triangle and carpet have inner edges. We construct trial states for both inner and outer edge states. We find that oscillations of the entropy as a function of particle number are present for the T-fractal, but not for the Sierpinski carpet. Finally, we observe deviations from the area law for several different bipartitions on the Sierpinski triangle.
\end{abstract}

\section{Introduction}

When a two-dimensional electron gas is placed in a strong magnetic field and at very low temperatures, the fractional quantum Hall (FQH) effect can appear \cite{tsui1982two}. The Hamiltonian describing this effect, which is basically the Coulomb interaction projected to the lowest Landau level, is macroscopically degenerate and extremely hard to diagonalize, except for a small number of particles. In 1983, R. B. Laughlin proposed an ansatz \cite{laughlin1983anomalous} for the ground state wavefunction of the FQH effect at odd-denominator filling factors. The so-called Laughlin wavefunction predicted the existence of quasihole excitations with fractional electric charge. It was shown numerically to have more than 99 percent overlap with the exact ground state of the Coulomb repulsion for small systems \cite{laughlin1983anomalous}. 

In 1987, Kalmeyer and Laughlin showed that the ground state of a frustrated Heisenberg antiferromagnet on a triangular lattice is well described by a Laughlin wave function for bosons in which the particle-coordinates were restricted to be the lattice coordinates \cite{kalmeyer1987equivalence}. This so-called Kalmeyer-Laughlin wave function showed that FQH physics can be found in settings different from where it was originally discovered. Later, several models describing FQH physics in lattices were proposed \cite{bergholtz2013,parameswaran2013}.

The work on FQH lattice models is, in part, motivated by the developments in quantum engineering, e.g.\ in ultracold atoms. Such systems also allow for going beyond periodic lattices, which opens several possibilities. With Rydberg atoms in optical tweezers arbitrary atomic arrangements can be obtained \cite{barredo2016,barredo2018}, and there are also ideas for implementing artificial gauge fields in these systems \cite{lienhard2020,Weber2022,Wu2022}, which is a starting point for obtaining quantum Hall physics.

Fractals are an interesting example of nonperiodic systems, because their Hausdorff dimension can be noninteger, and hence one can use the spatial dimension as a parameter to change the properties of the physical system. The quantum Hall effect was originally studied in two dimensions, but it has turned out that it can also appear in fractal dimensions. Examples of integer quantum Hall models in fractals \cite{Brzezinska2018,shriya2019,iliasov2020,Fremling2020,Fischer2021,Manna2022,ivaki2022} already show features that are not present for periodic lattices in two dimensions, such as inner edge states. FQH physics has also been investigated on fractal lattices \cite{manna2020anyons,LaughlinFractal,zhu2022,jaworowski2023}, and one of the interesting observations here is that the entropy can scale differently with subsystem size than it does in two dimensions.   

Trial states, such as the Laughlin state, are important tools to gain insight into the FQH effect. Moore and Read showed that the Laughlin state can be expressed as a correlation function in conformal field theory  \cite{moore1991nonabelions}, and this construction was later modified to obtain lattice versions of the Laughlin state that share the same topological features as the continuum Laughlin state for several different lattices and parameters \cite{nielsen2012laughlin,tu2014lattice}.
The construction can also be applied to obtain Laughlin states on fractal lattices as long as the fractal lattices can be embedded in the two-dimensional plane. This again leads to topological systems in many cases, allowing quasiholes to be added and investigated \cite{manna2020anyons}. The states were studied further on the Sierpinski triangle in \cite{LaughlinFractal}, where it was shown numerically that the entanglement entropy of the states does not follow an area law. It was also found that the density of particles on the Sierpinski triangle is different from the two-dimensional case.

In this paper, we take a deeper look into the properties of Laughlin states on different fractal lattices. In addition to the Sierpinski triangle, we also consider the Sierpinski carpet and the T-fractal. These fractals differ both with respect to Hausdorff dimension and ramification number. The considered fractal lattices and the lattice Laughlin wavefunction are described in section \ref{sec:wavefunction}. Since fractal lattices do not have a periodic bulk, the density of particles generally varies across the fractal lattice as observed for the Sierpinski triangle in \cite{LaughlinFractal}. In section \ref{sec:density}, we study how the density variations across the fractal lattice depend on the number of particles for the three fractals. We also find that the variations go to zero at half-filling of the lattice, which is due to particle-hole symmetry at half-filling. In section \ref{sec:anyons}, we introduce both quasiholes and quasiparticles into the lattice Laughlin state. We investigate how the presence of the anyons modify the density, and we investigate how the size of the anyons depend on their position in the fractal lattice. In section \ref{sec:edgestates}, we construct trial states for edge states on both the inner and outer edges of the Sierpinski triangle and the Sierpinski carpet. In section \ref{sec:SvM}, we compute the behaviour of the entropy as a function of particle number. It was found previously \cite{LaughlinFractal} that the entropy shows oscillations with a fixed period as a function of particle number for the Sierpinksi triangle, but not for the square lattice. Here, we find that oscillations also occur for the T-fractal, but not for the Sierpinski carpet. This suggests that the ramification number could play a role for the presence or absence of the oscillations. In section \ref{sec:scaling}, we investigate how the entropy scales with subsystem size for different bipartitions of the Sierpinski triangle. We find that different bipartitions give rise to different behaviors violating the area law. In section \ref{sec:conclusion}, we summarize the conclusions. In \ref{sec:appendix}, we describe analytically how the density and entanglement entropy transform under the particle-hole transformation.

\section{The lattice Laughlin wavefunction}\label{sec:wavefunction}

The lattice Laughlin states considered in \cite{nielsen2012laughlin,tu2014lattice} have the following form
\begin{eqnarray}\label{laughlin}
&|\psi\rangle=\sum_{n_1,\ldots,n_N} \psi(n_1,\ldots,n_N)
|n_1,\ldots,n_N\rangle,\\
&\psi(n_1,\ldots,n_N) =
C_M\delta_{n,M} \, \prod_{i<j} (z_i-z_j)^{q n_i n_j}
\prod_{k\neq l} (z_k-z_l)^{-\eta n_l},\nonumber
\end{eqnarray}
where the $z_i$ are the coordinates of the lattice sites written as complex numbers and $n_{i}$ are their occupation numbers, which are either $0$ or $1$, since the particles are fermions for odd $q$ and hardcore bosons for even $q$. The positive integer $q=1/\nu$ is the flux per particle. The Kronecker delta 
\begin{equation}
\delta_{n,M}\equiv \delta_{\sum_i n_i,M}=\left\{\begin{array}{ll} 
1 \textrm{ for }\sum_i n_i=M\\
0 \textrm{ otherwise}\end{array}\right.
\end{equation}
ensures that $\sum_{i} n_{i}=M$ satisfies $qM=N\eta$ for all nonzero terms of the wavefunction, where $M$ is the total number of particles and $N$ is the total number of lattice sites. Since $qM$ is the total number of fluxes, $\eta$ can be interpreted as the flux per lattice site. The constant of proportionality $C_M$ is chosen such that the wavefunction $|\psi \rangle$ is normalized. The last factor in (\ref{laughlin}) differs from the Gaussian factor that appears in the original Laughlin state, because the background charge is here placed on the lattice sites only. This point turns out to be important to obtain the desired physics on fractal lattices \cite{manna2020anyons}. An exact parent Hamiltonian for (\ref{laughlin}) can be found in \cite{tu2014lattice,manna2020anyons}, and \cite{jaworowski2023} made a systematic search for simpler, approximate parent Hamiltonians.

In the following sections, we study the properties of the wavefunction (1) when the $z_i$ are given by the coordinates of a finite generation fractal lattice or a square lattice. Specifically, we consider the Sierpinski triangle depicted in figure \ref{fig:densfrac}(a) with $N=3^g$ sites, the Sierpinski carpet in figure \ref{fig:densfrac}(d) with $N=8^g$ sites, and the T-fractal in figure \ref{fig:densfrac}(g) with $N=5^g$ sites, where $g$ is the generation. The infinite version of these fractals have Hausdorff dimensions $\ln(3)/\ln(2)\approx1.58$, $\ln(8)/\ln(3)\approx1.89$, and $\ln(5)/\ln(3)\approx1.46$, respectively. They also differ in ramification number \cite{balankin2019957}. If we imagine all the nearest-neighbor sites being linked by a bond, an arbitrarily large piece of the T-fractal can be isolated by cutting one bond, while that number is two for the Sierpinski triangle and infinite for the Sierpinski carpet.

\section{Particle density}\label{sec:density}

We first consider the particle density on the lattice sites, i.e.\ $\langle  \psi |\hat{n}_i|\psi \rangle$, where $\hat{n}_i=\hat{c}_i^{\dagger}\hat{c}_i$ is the number operator and $\hat{c}_i$ is the operator that annihilates a particle on the $i$th lattice site. This quantity can be computed via Monte Carlo simulations of the expression
\begin{equation}
\langle \hat{n}_i\rangle=\sum_{n_1,\ldots,n_N} n_i |\psi(n_1,\ldots,n_N)|^2.
\end{equation}

For Laughlin states on two-dimensional, periodic lattices, we expect the expectation value of the number operator to be uniform in the bulk of the system (see, e.g., \cite{LaughlinFractal} for computations on square and triangular lattices). For fractal structures, however, there is not a distinct edge or bulk. The density pattern is, hence, not obvious a priori. In \cite{LaughlinFractal}, the particle density of the Laughlin state on the $g=5$ Sierpinski triangle for a fixed particle number ($M=27$) was computed and compared to the density on a triangular lattice. While the density in the bulk of the triangular lattice is constant, there are fluctuations around the mean value near the edge. On the fractal lattice, the density fluctuates from site to site, roughly imitating the edge density pattern of the triangular lattice. It was also found that the density on the corner sites of the fractal differs significantly from the density on the other sites of the fractal lattice.

Here, we take this analysis further by studying the densities on the Sierpinski triangle as a function of $M$, as opposed to studying the spatial pattern for a fixed value of $M$. Moreover, we study the density using this approach also for the Sierpinski carpet and the T-fractal. At zero and complete filling the densities are trivially constant across the lattice, so we will not discuss those cases in the following. For almost all other values of $M$, we find that there is a spatial variation, but at exactly half filling, the variance goes to zero. The plots also have a symmetry around the half-filling point, which appears in all the studied fractals at all values of $q$. Both of these observations can be explained by studying the particle-hole transformation of the lattice Laughlin state as noted in \cite{LaughlinLattice} and discussed in detail in \ref{sec:appendix} below. We also see that the corner densities behave differently from the rest of the lattice, except near and at half-filling. We propose that the density structure can be used as a tool to define and distinguish a bulk and edge for a fractal where, a priori, there is no clear distinction between them.
 
\begin{figure*}
     \raisebox{34mm}{(a)}
     \includegraphics[width=0.25\linewidth]{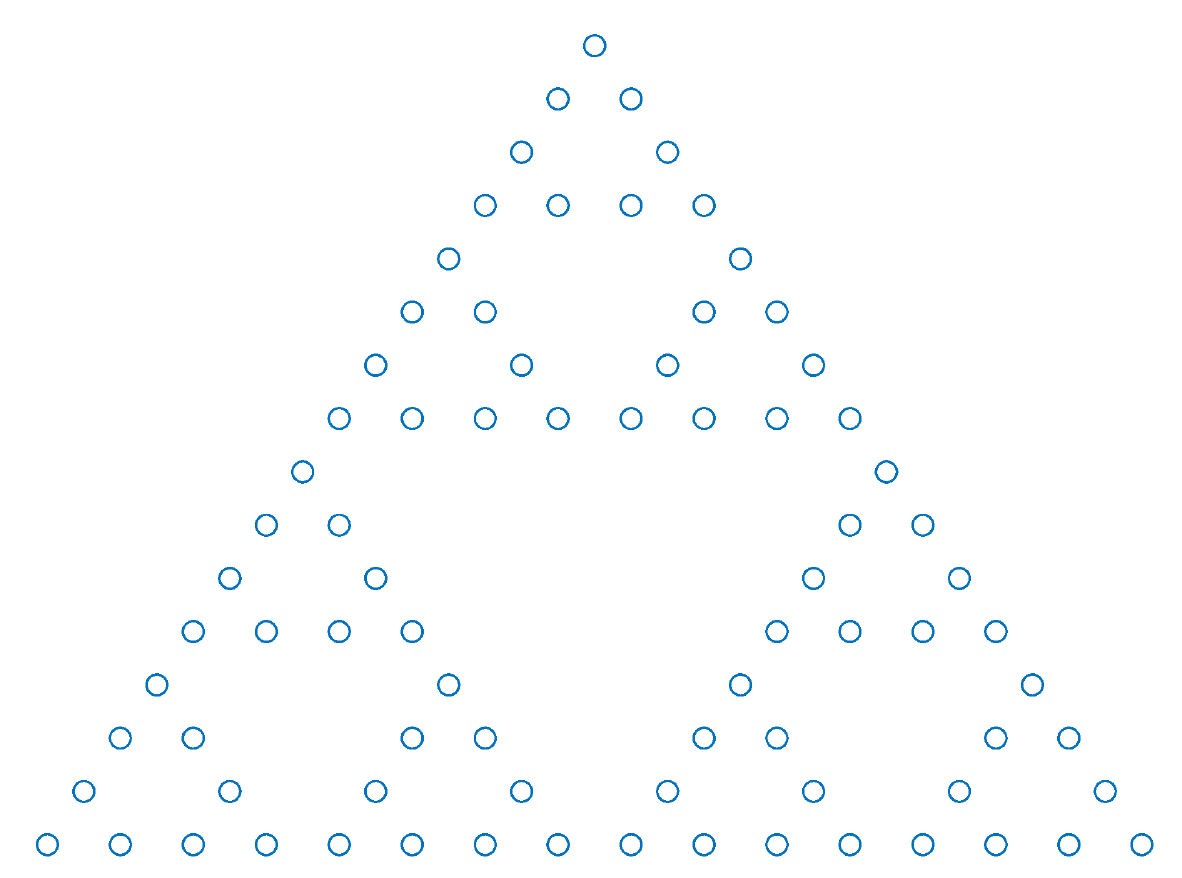}
     \raisebox{34mm}{(b)}
     \includegraphics[width=0.3\linewidth]{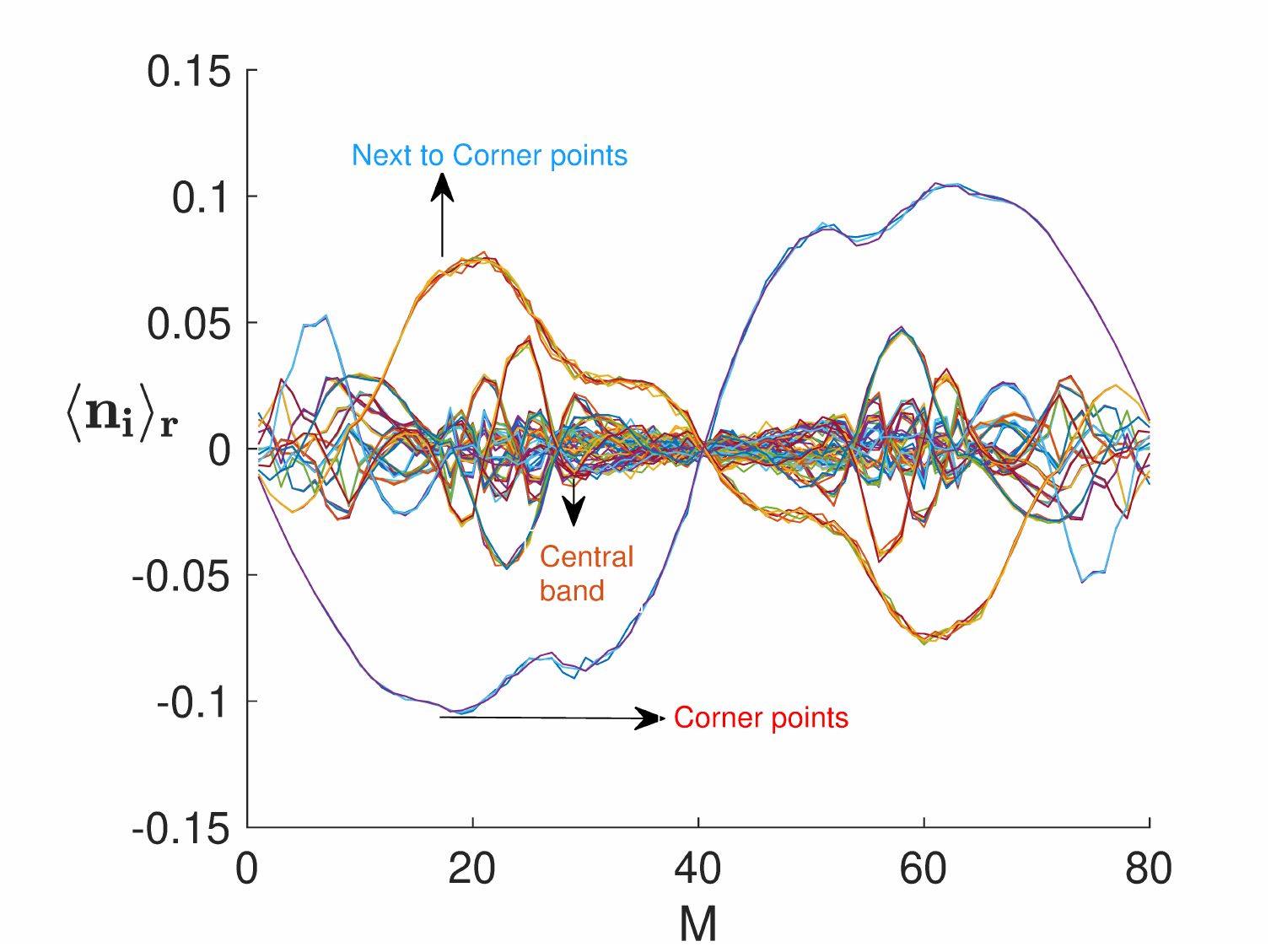}
     \raisebox{34mm}{(c)}
     \includegraphics[width=0.3\linewidth]{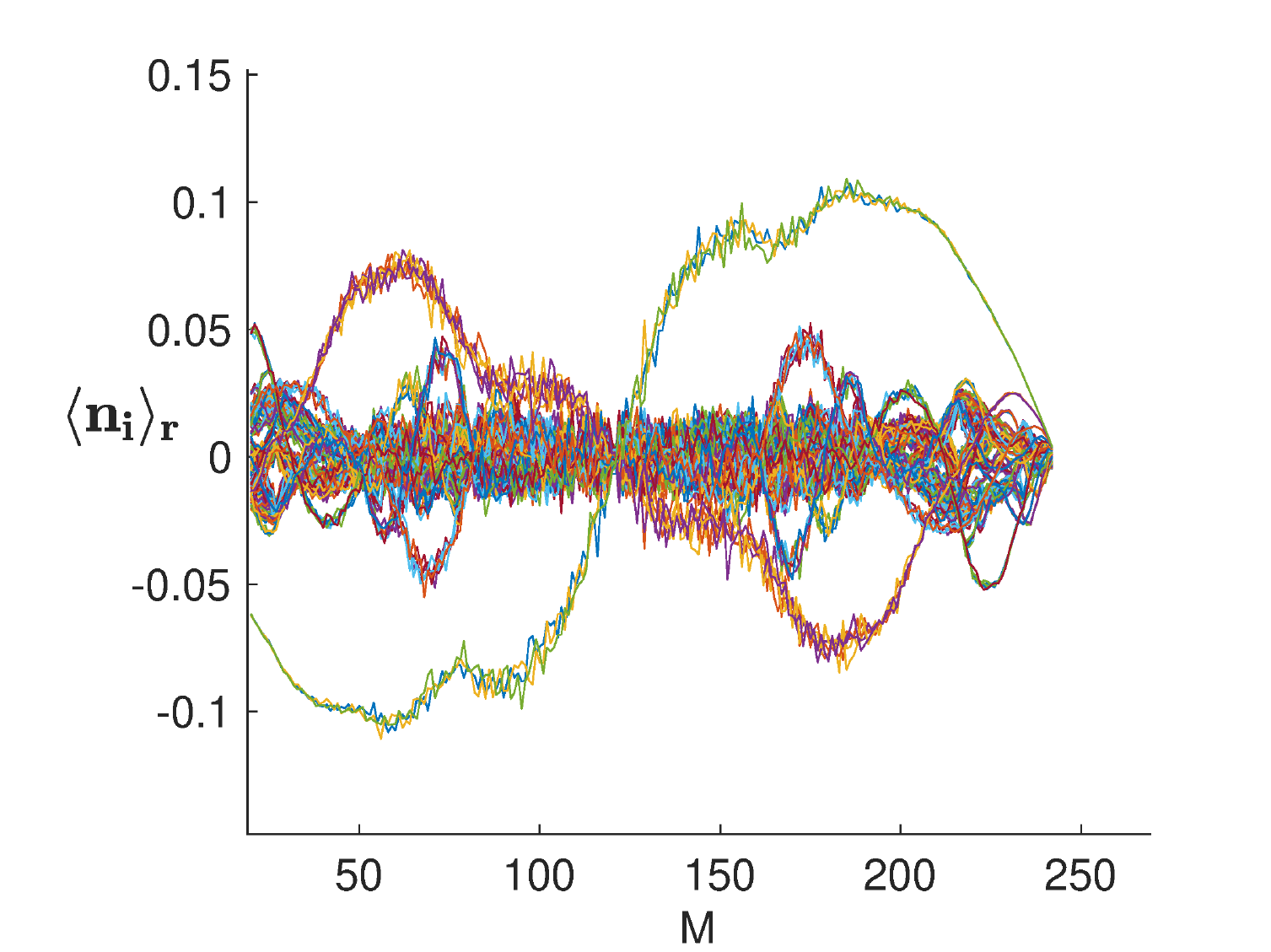}\\
     \raisebox{34mm}{(d)}
     \includegraphics[width=0.25\linewidth]{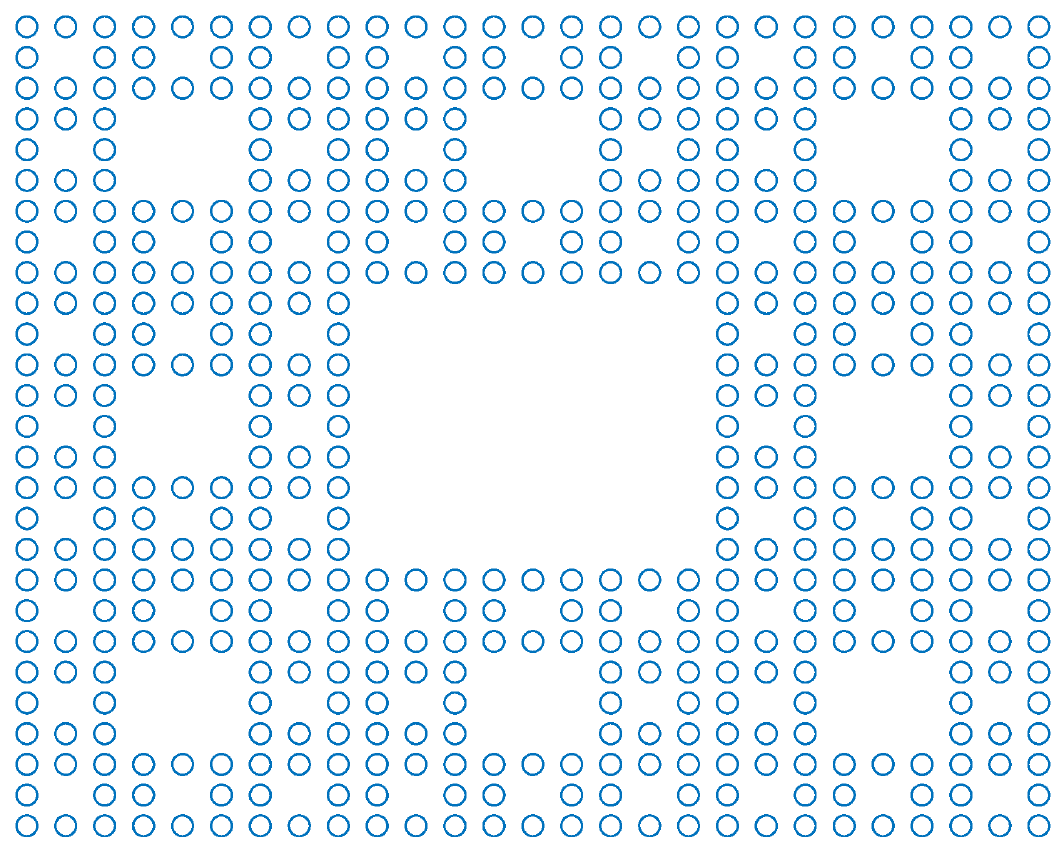}
     \raisebox{34mm}{(e)}
     \includegraphics[width=0.3\linewidth]{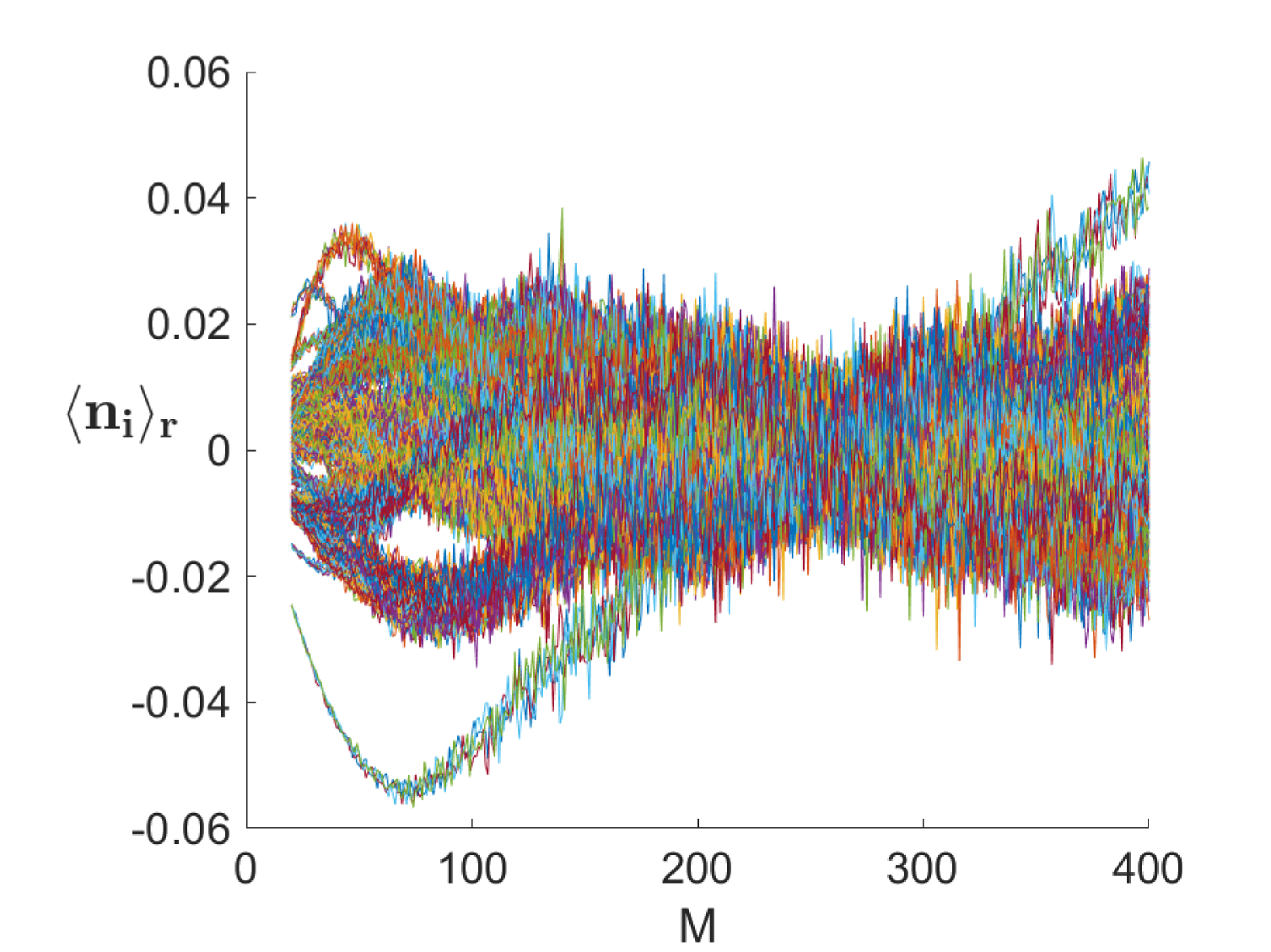}
     \raisebox{34mm}{(f)}
     \includegraphics[width=0.3\linewidth]{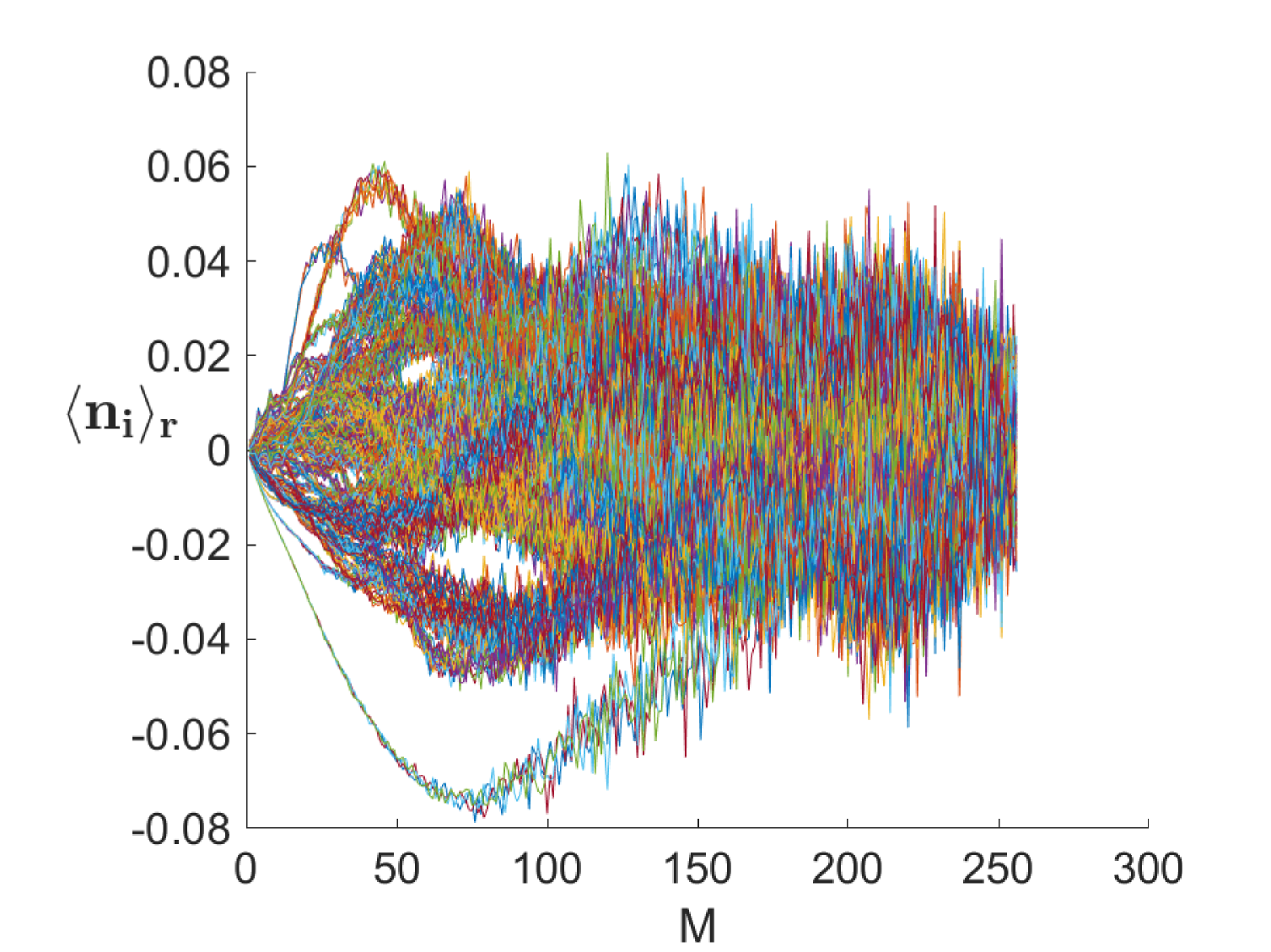}\\
     \raisebox{34mm}{(g)}
     \includegraphics[width=0.25\linewidth]{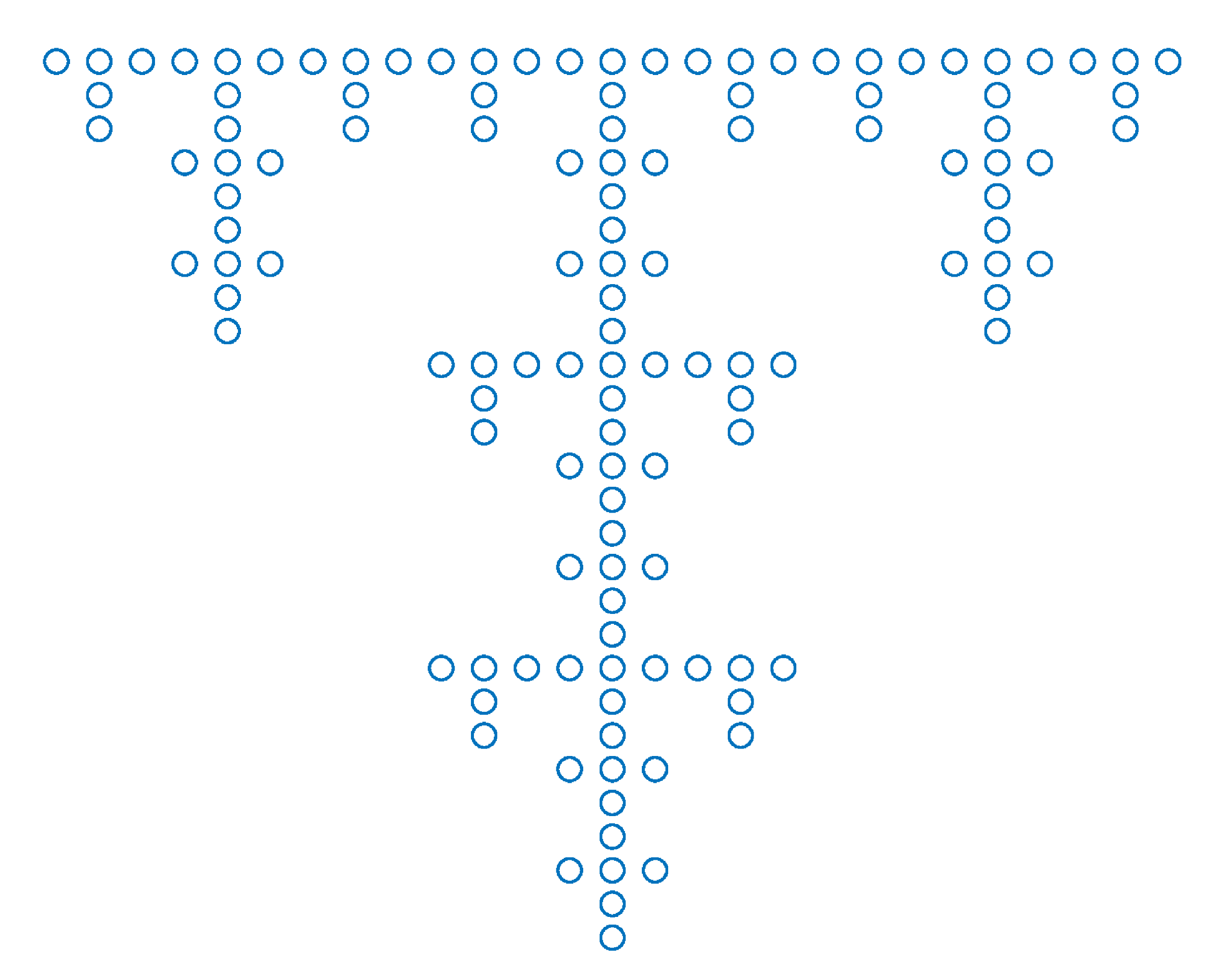}
     \raisebox{34mm}{(h)}
     \includegraphics[width=0.3\linewidth]{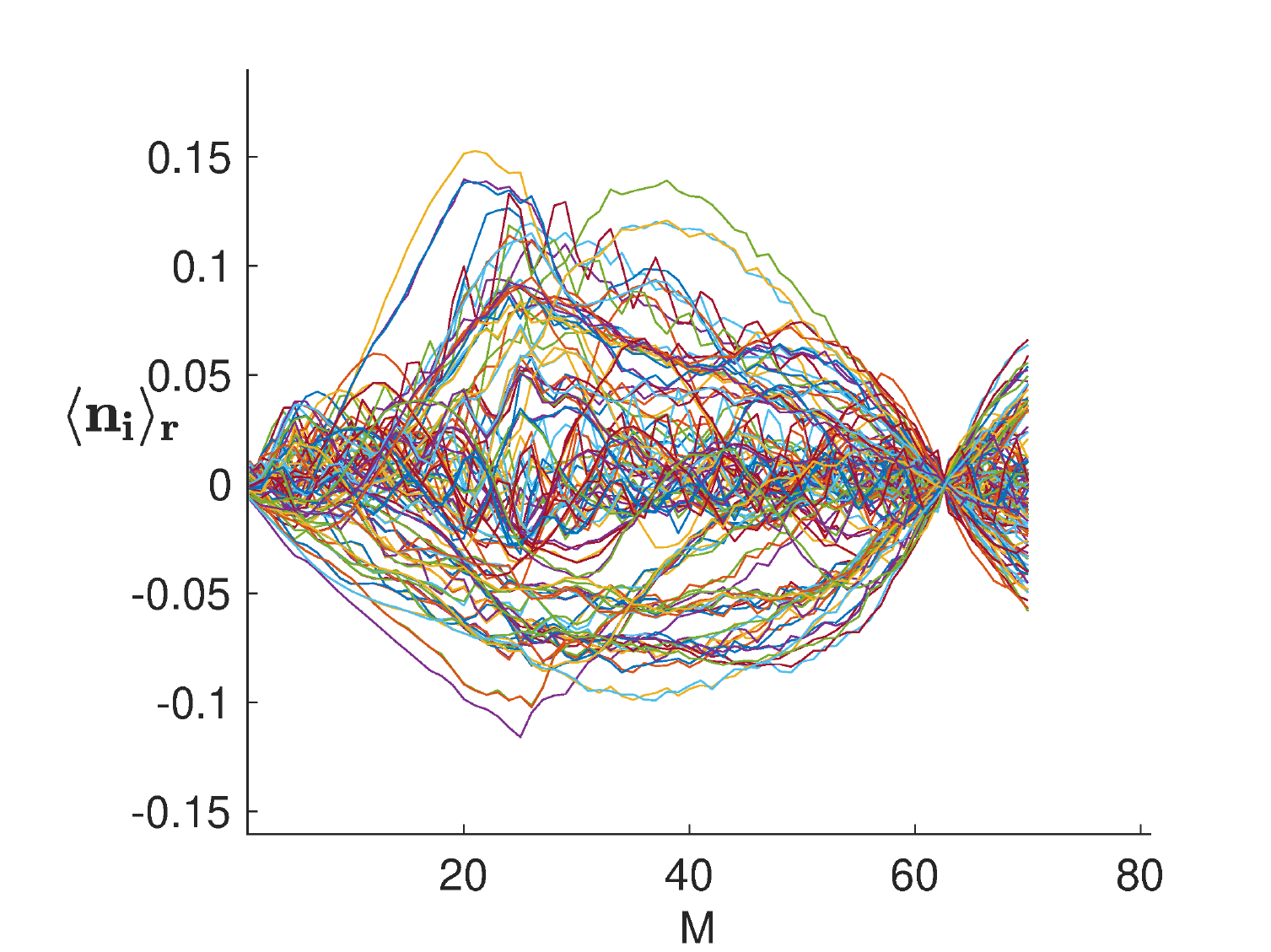}
     \raisebox{34mm}{(i)}
     \includegraphics[width=0.3\linewidth]{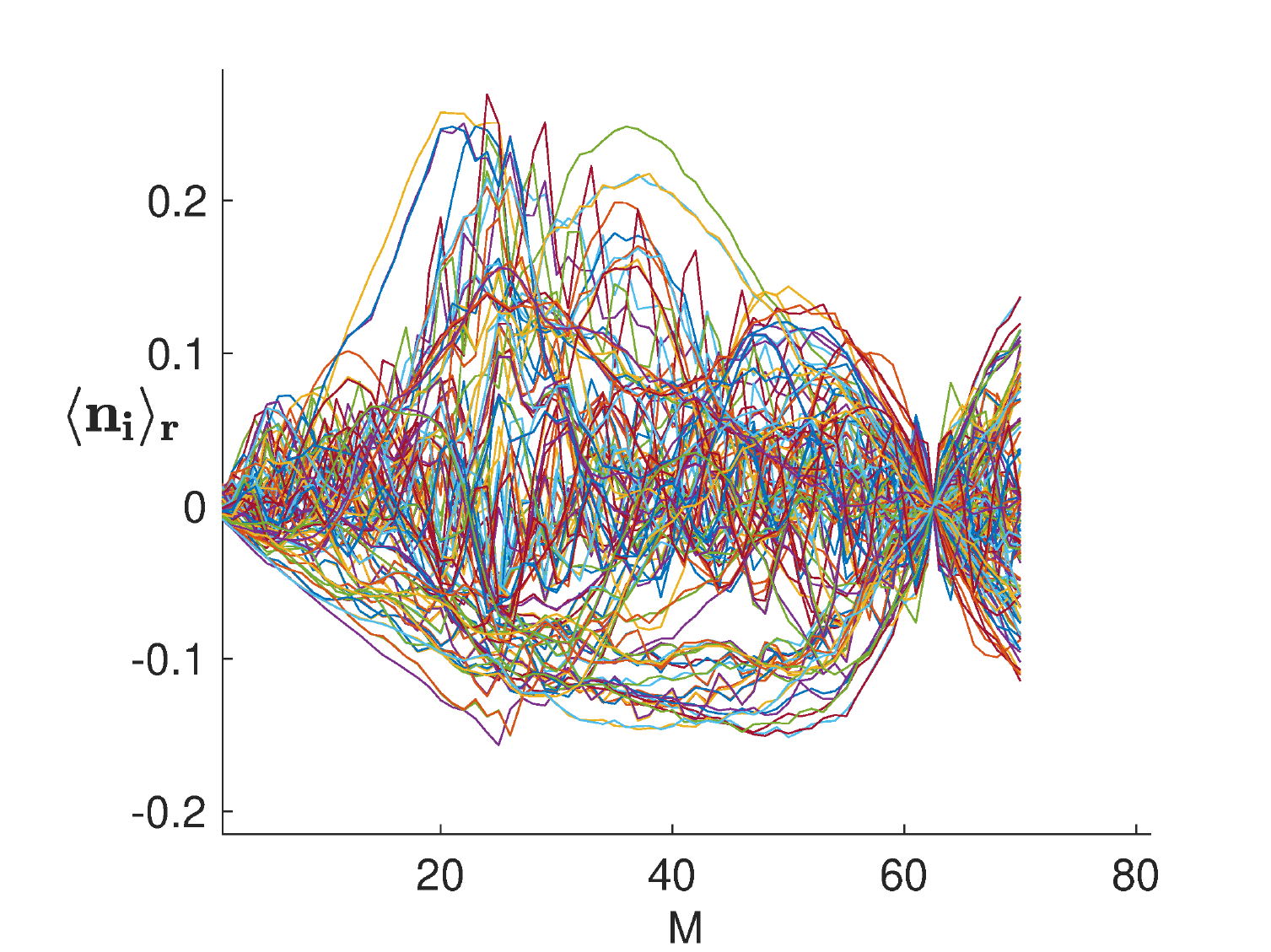}
     \caption{(a) The $g=4$ Sierpinski triangle with $N=81$ sites. (b) The particle densities of all 81 sites, plotted against particle number $M$, for the $q=3$ lattice Laughlin state. The densities $\langle n_{i} \rangle_{r}$ are obtained by subtracting the mean value $M/N$, i.e., $\langle n_{i} \rangle_{r} = \langle n_{i} \rangle - M/N $. (c) The particle densities of all $243$ sites of a $g=5$ Sierpinski triangle for the $q=3$ Laughlin state. (d) The $g=3$ Sierpinski carpet with $N=512$ sites. The particle densities of all 512 sites are shown for (e) $q=2$ and (f) $q=3$. (g) The $g=3$ T-fractal with $N=125$ sites. The particle densities for all $125$ sites are shown for (h) $q=2$ and (i) $q=3$}
      \label{fig:densfrac}
\end{figure*}

Figure \ref{fig:densfrac}(b) shows $ \langle n_i \rangle_r= \langle n_i \rangle - M/N$ of a generation $g=4$ Sierpinski triangle for all values of the number of particles $M$ in the range [1,80]. The data roughly form a thick central band and two sets of lines (blue and reddish) that are quite far from the central band for most values of $M$. The blue line is composed of the three sites at the corners of the fractal, which all have the same density because of the $C_{3}$ symmetry of the Sierpinski triangle. The reddish line is composed of the six sites that are neighbors of the three corner sites.

There is a significant dispersion within the central band at lower values of $M$. At $M=27$, however, which corresponds to the case of $\eta=qM/N=1$, the dispersion decreases somewhat. After this the dispersion rises slightly and briefly until $M=30$, at which point it starts falling monotonically. At $M=40$ and $M=41$ (corresponding to almost half filling) it is close to zero. As a whole, the densities reflect the particle-hole symmetry. This means that the plot of $\langle n_i \rangle_r$ as a function of $M$ is unchanged by a rotation by $\pi$ around the point ($\langle n_i \rangle_r=0,M=N/2$). In other words, $\langle n_i \rangle_{r,M}=-\langle n_i \rangle_{r,N-M}$, where $\langle n_i \rangle_{r,M}$ is $\langle n_i \rangle_{r}$ for the state with $M$ particles. This behaviour of the densities is derived analytically in \ref{sec:appendix}.

In figure \ref{fig:densfrac}(c), the densities are plotted for the next generation of the fractal with $243$ sites and $q=3$. We see that the structure of a central band surrounded by deviant lines corresponding to corner sites and their immediate neighbours becomes even more apparent. Again we see a pinching of the densities at $\eta=1$, i.e.\ $M=81$, as well as close to half-filling ($M=120,121$). At $\eta=1$, only the central band seems to pinch while the corner (and next to corner) lines remain separated, but at half-filling, even the densities at the corner sites go to zero. 

We next consider the Sierpinski carpet in figure \ref{fig:densfrac}(d). The densities are shown in figure \ref{fig:densfrac}(e-f). The values for larger fillings can be obtained utilizing the symmetry discussed above. We see a central band which is initially separated from a set of lines below it. The points below correspond to the four corner sites. At around $M=150$, they merge with the central band. For the case of $q=2$, this happens at around $M=175$. For both values of $q$, $\langle n_i \rangle_r$ goes to zero at half filling (i.e.\ $M=256$) for all lattice sites. This is not immediately visible in the figure, however, due to the large number of points plotted. 

In figure \ref{fig:densfrac}(h-i), we plot the particle densities for the T-fractal. Compared to the Sierpinski triangle and carpet, there are larger variations in the densities with several sites having $|\langle n_i \rangle_r|>0.05$. At half-filling, the densities converge to zero as for the other fractal lattices.

\section{Quasiholes and quasiparticles} \label{sec:anyons}

In this section, we investigate quasiholes and quasiparticle in the Laughlin states. We first give examples of density profiles of the anyons, and we then study the size of the anyons as a function of position on the fractal lattices. Building on \cite{laughlin1983anomalous}, it was shown in \cite{Nielsen_2018} that states describing quasiholes and quasiparticles on periodic lattices with open boundary conditions can be obtained by modifying the lattice Laughlin state (\ref{laughlin}) into    
\begin{eqnarray}\label{lanyon}
&|\psi_{\vec{p},\vec{w}}\rangle=\sum_{n_1,\ldots,n_N} \psi_{\vec{p},\vec{w}}(n_1,\ldots,n_N)
|n_1,\ldots,n_N\rangle,\\
&\psi_{\vec{p},\vec{w}}(n_1,\ldots,n_N) \propto
\delta_{n,M} \, 
\prod_{i,j} (w_i-z_j)^{p_i n_j}
\prod_{i<j} (z_i-z_j)^{q n_i n_j}
\prod_{k\neq l} (z_k-z_l)^{-\eta n_l},\nonumber
\end{eqnarray}
where $w_i$ is the coordinate of the $i$th anyon, and $p_i=+1$ ($p_i=-1$) if the $i$th anyon is a quasihole (quasiparticle). Also, 
\begin{equation}
\eta=\left(qM +\sum_i p_i\right)/N.
\end{equation}
The anyons are extended objects that appear as local variations in the density. The $w_i$ can be anywhere in the complex plane, and the density variations appear on the lattice sites close to $w_i$, i.e.\ the anyon lives on the lattice sites even if $w_i$ does not coincide with lattice sites. The sum of the density variations over a region that is large enough to contain the anyon is $-p_i/q$, so for $q=2$, for instance, a quasihole gives rise to half a particle missing on average in a local region, while a quasiparticle gives rise to half a particle extra on average in a local region. One can show \cite{Nielsen_2018} that the braiding properties are as expected for anyons in systems with Laughlin type topology, as long as (\ref{lanyon}) produces the correct local density variations and the local density variations do not overlap during the braiding process. It is hence sufficient to study the density variations to see the topology.

This construction also applies to fractal lattices, but for each lattice considered one needs to check numerically whether the state still describes anyons, i.e.\ whether (\ref{lanyon}) produces the correct changes in density compared to (\ref{laughlin}). This question was studied for quasiholes in \cite{manna2020anyons}. Here, we extend that analysis by also considering quasiparticles, by considering further fractals (specifically the Sierpinski carpet and the T-fractal), and by quantifying the size of the anyons as a function of position on the fractal lattice.

We shall here consider states with one quasihole and one quasiparticle, such that $\sum_i p_i=0$, and we shall take $q=2$ throughout. To limit the number of possible positions of the anyons, we shall take all $w_i$ to be on lattice sites, but the case where the $w_i$ are not on lattice sites can be studied with the same approach. To quantify the density variations, we define
\begin{equation}\label{rhoi}
\rho_i=\langle \psi_{\vec{p},\vec{w}}| \hat{n}_i|\psi_{\vec{p},\vec{w}}\rangle-\langle \psi|\hat{n}_i|\psi\rangle.
\end{equation}
We shall also consider 
\begin{equation}\label{Dr}
D_r=\sum_i \rho_i \theta(r-|z_i-w_k|)
\end{equation}
for the $k$th anyon, i.e.\ the sum of $\rho_i$ over a circular region around $w_k$. The Heaviside step function $\theta$ selects the $i$ for which the distance between $z_i$ and $w_k$ is less than $r$. For a system with properly screened anyons, $D_r$ for the $k$th anyon approaches $-p_k/q$ when $r$ is large, but still small compared to the distance to other anyons in the system. We compute $\rho_i$ and $D_r$ through Monte Carlo simulations of $\langle \psi_{\vec{p},\vec{w}}| \hat{n}_i|\psi_{\vec{p},\vec{w}}\rangle$ and $\langle \psi|\hat{n}_i|\psi\rangle$.

\begin{figure}
\includegraphics[width=\linewidth]{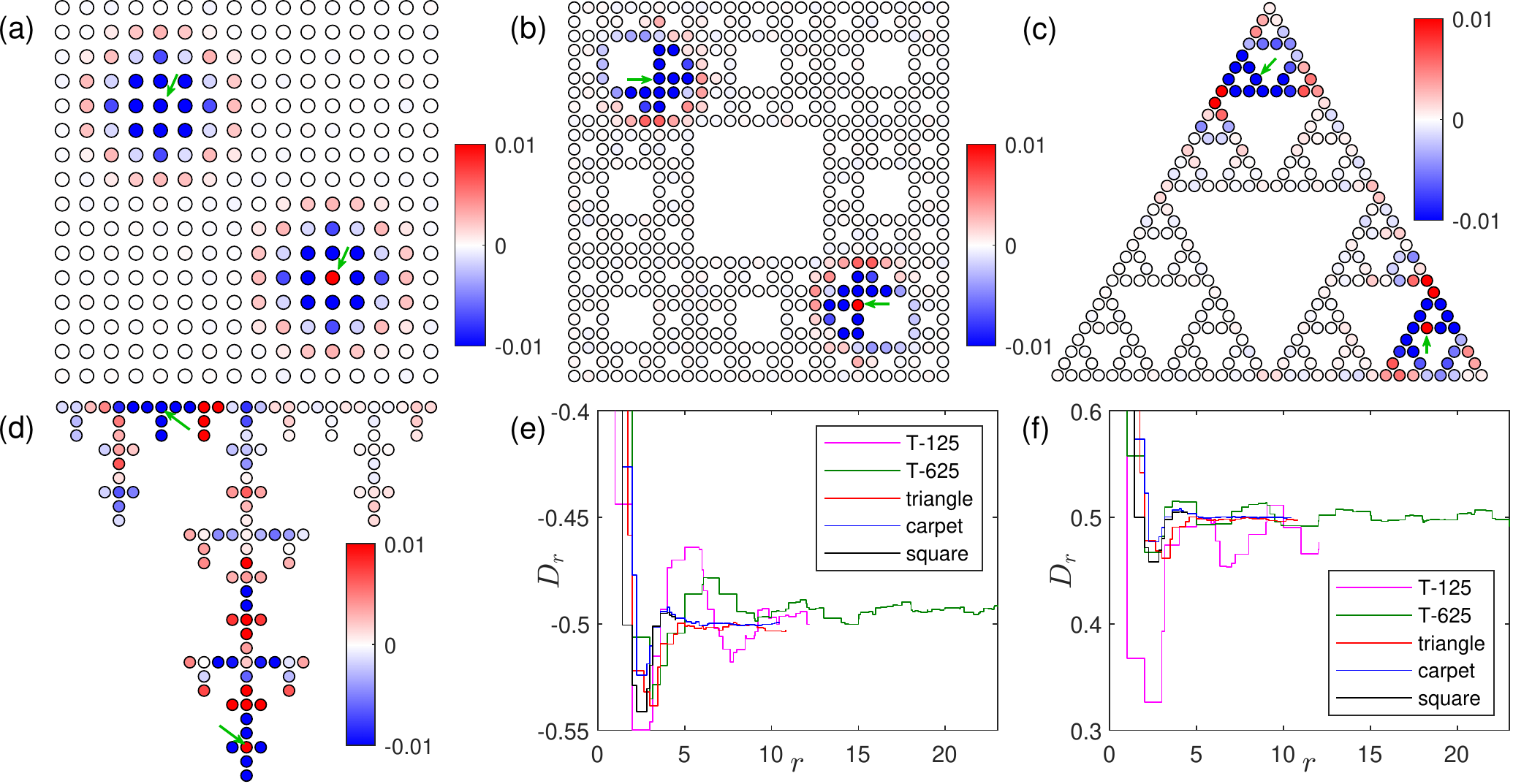}
\caption{A quasihole and a quasiparticle with the $w$-coordinates placed at the sites marked by the green arrows on (a) a square lattice with $256$ sites, (b) a Sierpinski carpet with $512$ sites, (c) a Sierpinski triangle with $243$ sites, and (d) a T-fractal with $125$ sites. The color of the lattice sites shows $\rho_i$ (see (\ref{rhoi})). To make finer details visible, the color bars have been saturated at $\pm0.01$, so values outside this range are shown as dark blue or dark red. The number of particles is chosen such that the lattice filling $M/N$ is as close to $1/9$ as possible for all lattices, and we take $q=2$. $D_r$ (see (\ref{Dr})) is shown as a function of the radius $r$ for (e) the quasihole and (f) the quasiparticle. The radius is measured in units of the distance between nearest neighbor sites. For the square lattice, the Sierpinski carpet, and the Sierpinski triangle, $D_r$ is seen to be close to $-1/2$ and $1/2$, respectively, for $r$ larger than about $5$, while there are larger fluctuations for the T-fractal. The fluctuations decrease in size, however, when the number of sites is increased from $125$ (T-125) to $625$ (T-625). The lines are terminated at half the distance between the anyons, except for the T-fractal with $625$ sites, for which half the distance is outside the interval shown.}\label{fig:Anyons}
\end{figure}

Results for the square lattice, the Sierpinski carpet, the Sierpinski triangle, and the T-fractal are given in figure \ref{fig:Anyons}. For all cases, we take $M/N$ as close to $1/9$ as possible to be able to compare the different lattices. For the square lattice, the Sierpinski carpet, and the Sierpinski triangle, well-separated quasiholes and quasiparticles are seen, and $D_r$ is close to $\pm1/2$ for $r$ larger than about $5$ measured in units of the lattice spacing. It is also seen that the quasihole and the quasiparticle have approximately the same size. For the T-fractal with $125$ sites the separation is less clear, and $D_r$ has larger variations for $r$ around 5 to 10 than for the other lattices. By increasing the system size to $625$ sites, we can increase the separation between $w_1$ and $w_2$, and we observe that $D_r$ is closer to $\pm1/2$ for large $r$ for this larger lattice. This suggests that we can have well-separated anyons in large enough T-fractals. It seems plausible that the larger size of the anyons in the T-fractal appears because several sites in the T-fractal only have two neighbors, which produces a less good screening.    

\begin{figure}
\includegraphics[width=\linewidth]{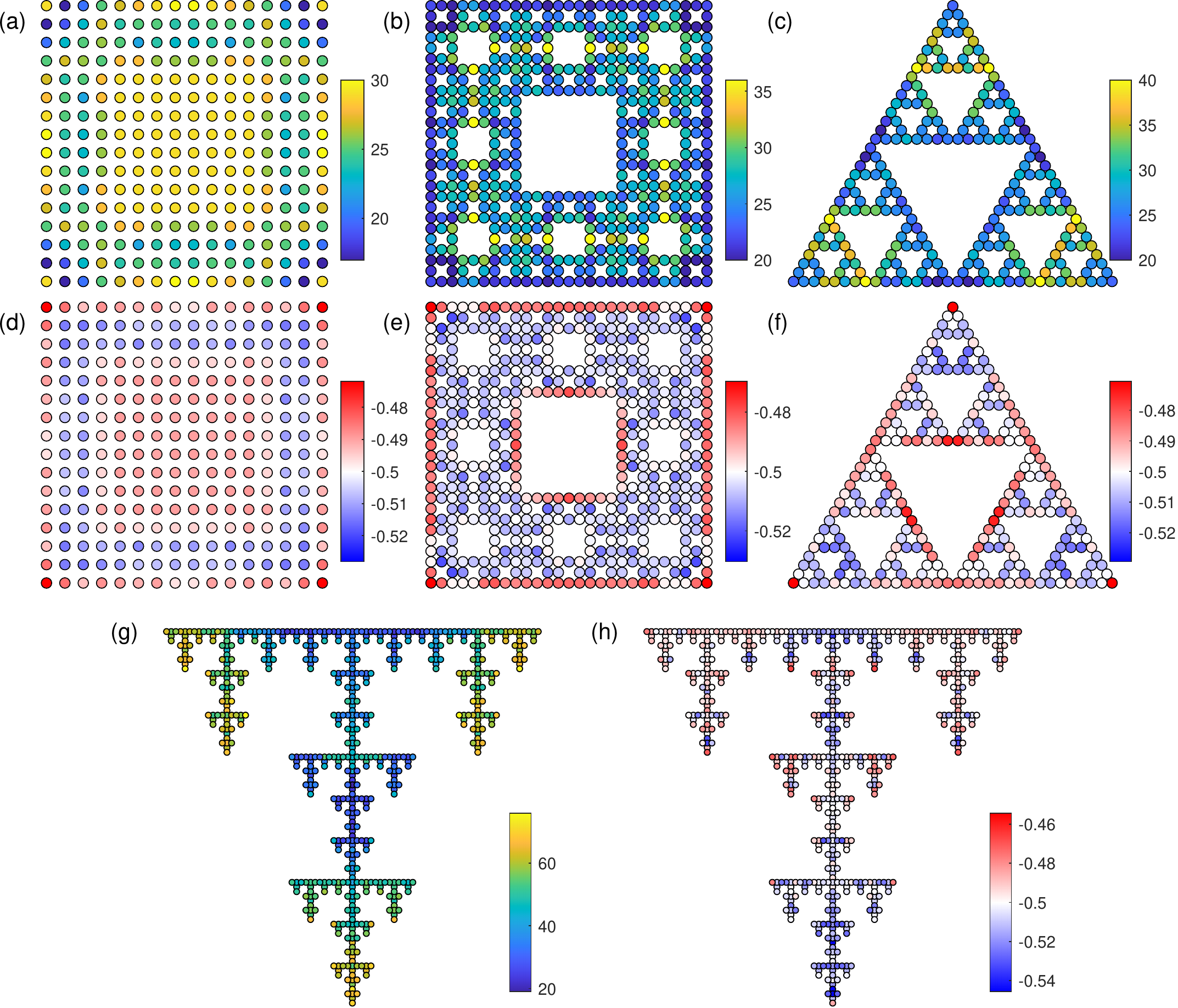}
\caption{Size of quasiholes. We consider $q=2$ and place the $w$-coordinate of a quasihole on lattice site $i$. The $w$-coordinate of a quasiparticle is placed on the lattice site, which is the longest distance away from site $i$ measured in the two-dimensional plane. We consider a disk around the quasihole with a radius that is half the distance between the anyons. To estimate the size of the quasihole, we show the number of sites within this disk for which $|\rho_i|>0.002$ in (a), (b), (c), and (g), and we show the sum of $\rho_i$ over the sites within the disk with $|\rho_i|>0.002$ in (d), (e), (f), and (h).}\label{fig:SizesP}
\end{figure}

\begin{figure}
\includegraphics[width=\linewidth]{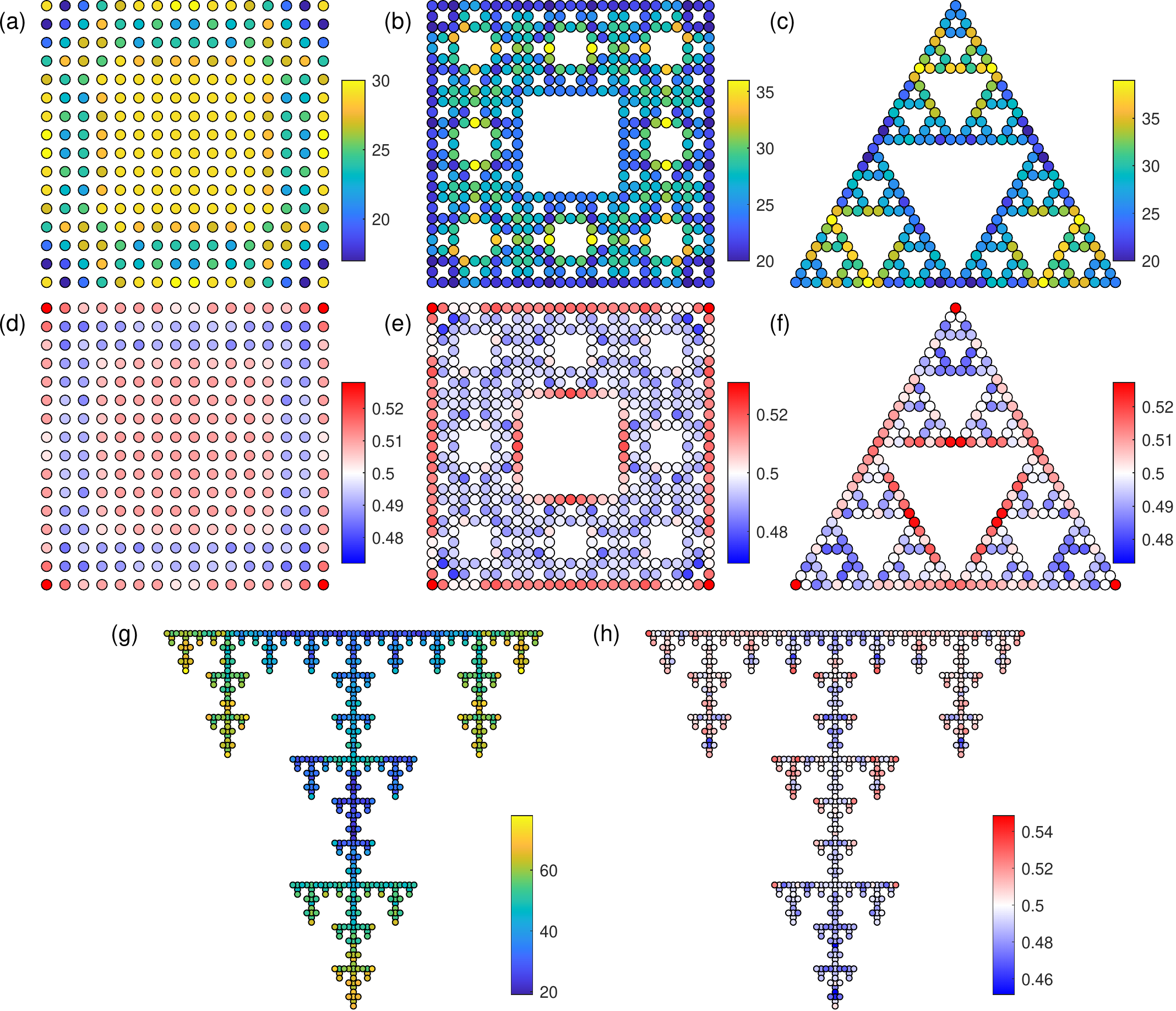}
\caption{Size of quasiparticles. The plotted quantities are the same as in figure \ref{fig:SizesP}, except that the $w$-coordinate of the quasiparticle is now placed on site $i$, while the $w$-coordinate of the quasihole is placed on the site furthest from site $i$.}\label{fig:SizesM}
\end{figure}

We next make a more systematic study of the size of the quasiholes and the quasiparticles as a function of position on the lattice. It is not straightforward to precisely define the size of an anyon, as the $\rho_i$ are only defined on lattice sites. Here, we quantify the size of an anyon as the number of lattice sites within a disk shaped region that have $|\rho_i|>0.002$. Note that this number is large compared to the Monte Carlo errors. The disk shaped region is centered at $w_k$ and its radius is half the distance to the nearest $w_j$ with $j\neq k$. We are, of course, interested in the situation, where the quasihole and the quasiparticle are well-separated, and we hence put the other anyon at the lattice site furthest away from $w_k$. The computed sizes are shown for the quasiholes in figure \ref{fig:SizesP} and for the quasiparticles in figure \ref{fig:SizesM}. We also show the sum of $\rho_i$ over the sites inside the disk shaped region with $|\rho_i|>0.002$. This number should be close to $-1/2$ for quasiholes and close the $+1/2$ for quasiparticles in order for the size estimates to be reasonable. 

Comparing the two figures, the results are seen to be quite similar for quasiholes and quasiparticles. For the square lattice, we observe that the anyon size is constant in the bulk of the system due to the periodicity, while the sizes vary on the edges. For the fractal lattices, the sizes vary more with position and depend on the lattice structure. For the Sierpinski carpet, the largest sizes appear for sites that are not on the outer edges and have only two nearest neighbors. For the Sierpinski triangle all sites except the three corner sites have three nearest neighbors, and here the largest sizes appear for positions that are not close to one of the corners and also not close to the central part of the fractal lattice. For the T-fractal on the other hand, the largest sizes appear for positions furthest to the left, right, or bottom of the fractal lattice. We also observe from the figure that the anyon sizes increase with decreasing Hausdorff dimension of the fractals. For a two-dimensional system, the linear size of a region scales as the square root of the number of sites in the region. For a one-dimensional system linear sizes instead scale linearly with the number of sites. For the considered fractals, the scaling is in between these two cases. The linear size of the anyons hence increase even faster with decreasing Hausdorff dimension than the sizes considered here, meaning that the lattice should be bigger to avoid overlap between anyons.

\section{Trial states for inner and outer edge states} \label{sec:edgestates}

In studies of chiral topological phases in noninteracting systems on fractal lattices, it has been observed that some of the states in the spectra are localized on inner edges of the lattices \cite{Brzezinska2018,Fischer2021}. One may therefore ask, if such inner edge states can also appear in interacting systems. The Laughlin state can be expressed as a correlator in conformal field theory, and for periodic systems with open boundaries, it has been found that trial edge states can be obtained by adding further operators into the correlator \cite{wen1992theory,dubail2012,estienne2013,herwerth2015}. A special case of this is to add fluxes. Here, we add fluxes to the lattice Laughlin state on the Sierpinski triangle and carpet and show that this gives rise to density modifications along the (inner) edges surrounding the added flux. Due to the screening properties of the states, the properties of the states away from the edges are not expected to be affected, and hence the states with added fluxes can be interpreted as (inner) edge states.

\begin{figure}
\includegraphics[width=\linewidth]{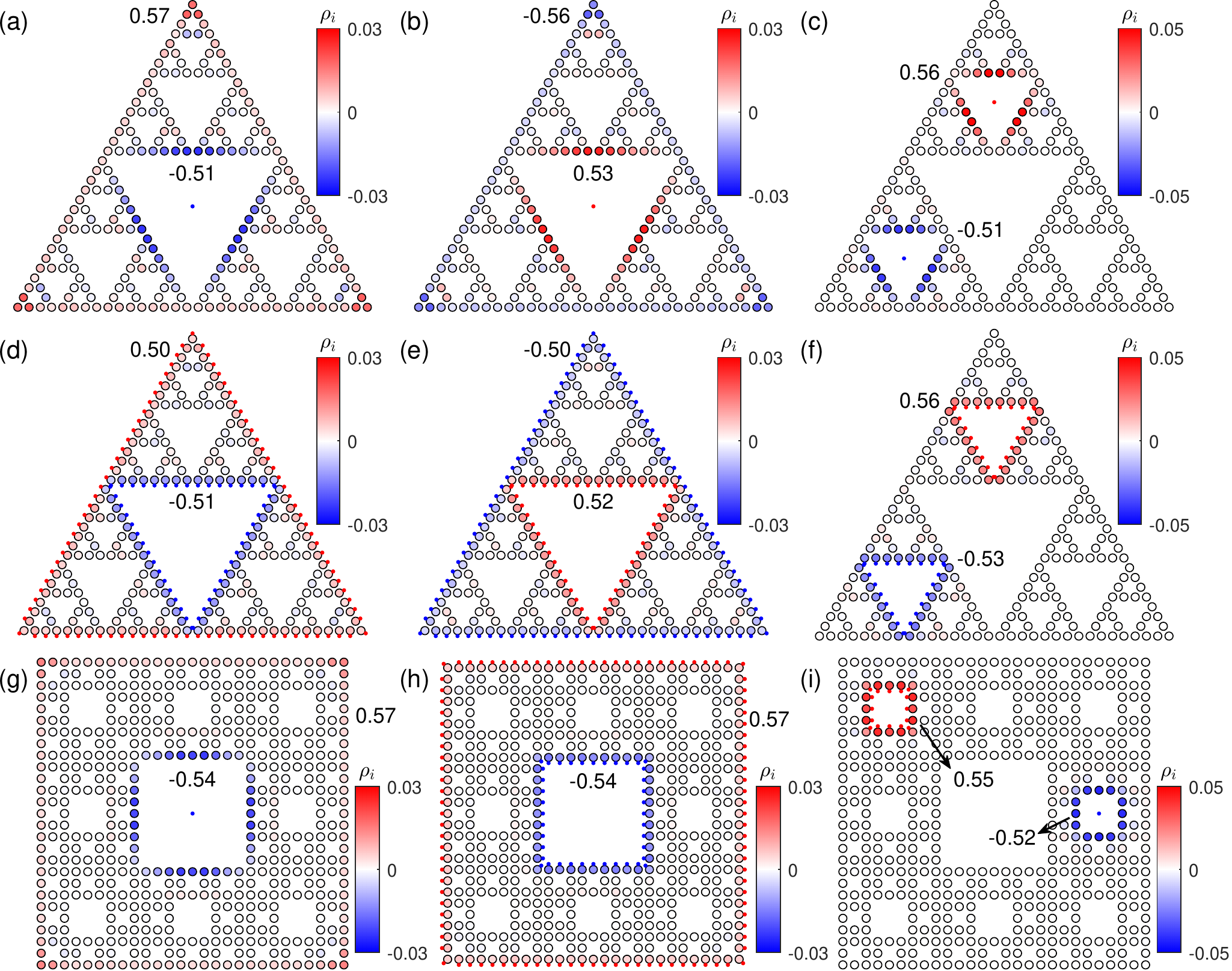}
\caption{Inner and outer edge states on the Sierpinski triangle and carpet produced by inserting fluxes in the Laughlin state at $q=2$. The color shows the density with the fluxes inserted minus the density when fluxes are not inserted. Color primarily on the edges hence signifies edge states. The blue and red dots are the positions of the positive and negative fluxes (in (a), (b), and (g) one flux is placed at infinity).}\label{fig:Edges}
\end{figure}

Our starting point is again the state (\ref{lanyon}). In this state, anyons are added to the lattice Laughlin state by adding extra fluxes at the positions $w_i$, which lead to the density modifications that are the anyons. Here, we instead add the fluxes in regions, where there are no lattice sites, as we expect this produces density variations on the edges closest to were the flux is added. We start by putting one flux at infinity and one flux at the center of the fractal lattice. We choose the fluxes to have opposite signs, i.e.\ $p_1=-p_2$. The resulting density modifications are shown in figure \ref{fig:Edges}(a,b,g) for the Sierpinski triangle and carpet. Alternatively, one can put both fluxes at the center of holes in the fractal lattice as illustrated in figure \ref{fig:Edges}(c). The figure shows that the fluxes indeed lead to density variations on the sites closest to the fluxes. When the fluxes are placed at the center of holes, inner edge states are formed on the sites surrounding the holes, and the fluxes at infinity produce density modifications on the outer edges of the fractal lattices. It is also seen in the figure that the produced density modifications are not uniform along the edges. This is particular clear for the outer edges, where the fluxes at infinity primarily lead to density modifications of the sites at the corners of the fractal lattices. Density variations that are more uniform along the (inner) edges can be obtained by splitting the fluxes into several pieces and placing them next to the sites forming the (inner) edges as illustrated in figures \ref{fig:Edges}(d,e,f,h,i). Computationally this is done by splitting each $w_i$ into $K$ positions $w_{i,k}$ and replacing the factors $\prod_{i} (w_i-z_j)^{p_i n_j}$ in (\ref{lanyon}) by $\prod_{i,k} (w_{i,k}-z_j)^{p_i n_j/K}$. The figures also display the sum of $\rho_i$ over the sites at the (inner) edges. Deviations of these numbers from $\pm1/2$ is a measure of to what extent the (inner) edges penetrate deeper into the lattice than the outermost layer of sites.

\section{Entropy as a function of particle number} \label{sec:SvM}

The entanglement entropy is a measure of the amount of entanglement between two parts of a quantum mechanical system. In \cite{LaughlinFractal}, it was found that the entanglement entropy shows oscillations as a function of the particle number $M$ on the Sierpinski triangle, while such oscillations are absent for the square lattice. Here, we find that such oscillations are also present for T-fractals, but not for the Sierpinski carpet. The origin of the oscillations is an open problem, but the results presented here suggest that the ramification number could play a role, as for the studied fractals, oscillations are seen only for the fractals with finite ramification number.

We consider the Renyi entropy $S_A$, which is defined as follows. We start from a pure state $|\psi \rangle$ and subdivide the system into two parts $A$ and $B$. The \textit{reduced} density matrix of subsystem $A$ reads $\rho_A=\mathrm{Tr}_B(|\psi\rangle\langle\psi|)$. Here, all the degrees of freedom of subsystem $B$ are traced over. Then the Renyi entanglement entropy of subsystem $A$ is defined as
\begin{equation}
S_A=-\mathrm{ln}[\mathrm{Tr}(\rho_A^2)].
\end{equation}
We consider the Renyi entropy rather than the von Neumann entropy, since the former can be computed efficiently with Monte Carlo simulations using the so-called replica trick \cite{cirac2010infinite,hastings2010measuring}. The trick is to simulate
\begin{equation}\label{replica}
\fl\rme^{-S_A}=
\sum_{n,m,n',m'}\delta_{(n',m),M}\delta_{(n,m'),M}\frac{\tilde{\psi}(n,m')\tilde{\psi}(n',m)}{\tilde{\psi}(n,m)\tilde{\psi}(n',m')}
|\psi(n,m)|^2 |\psi(n',m')|^2,
\end{equation}
which is an average of a quantity over the probability distribution $|\psi(n,m)|^2 |\psi(n',m')|^2$. Here, $n=\{n_1,n_2,\ldots,n_{N_A}\}$ is a basis for subsystem $A$, $m=\{n_{N_A+1},n_{N_A+2},\ldots,n_{N}\}$ is a basis for subsystem $B$, and $n'$ and $m'$ are independent copies of $n$ and $m$. We choose the numbering of the sites such that the sites in $A$ are number $1$ to $N_A$, and the sites in $B$ are number $N_A+1$ to $N$. The delta functions are defined as 
\begin{equation}
\delta_{(n',m),M}=\left\{\begin{array}{ll}
1 \textrm{ for }\sum_{i=1}^{N_A}n'_i+\sum_{i=N_A+1}^N n_i=M\\
0 \textrm{ otherwise}
\end{array}\right.,
\end{equation}
and 
\begin{equation}
\tilde{\psi}(n_1,\ldots,n_N)\equiv\prod_{i<j} (z_i-z_j)^{q n_i n_j}
\prod_{k\neq l} (z_k-z_l)^{-\eta n_l}. 
\end{equation}
Plots of $S_A$ versus $M$ for the state (\ref{laughlin}) have a mirror symmetry around $M=N/2$. This symmetry was observed and explained in \cite{LaughlinLattice}, and a detailed derivation is given in \ref{sec:appendix} below.

\begin{figure} 
    \includegraphics[width=0.32\linewidth]{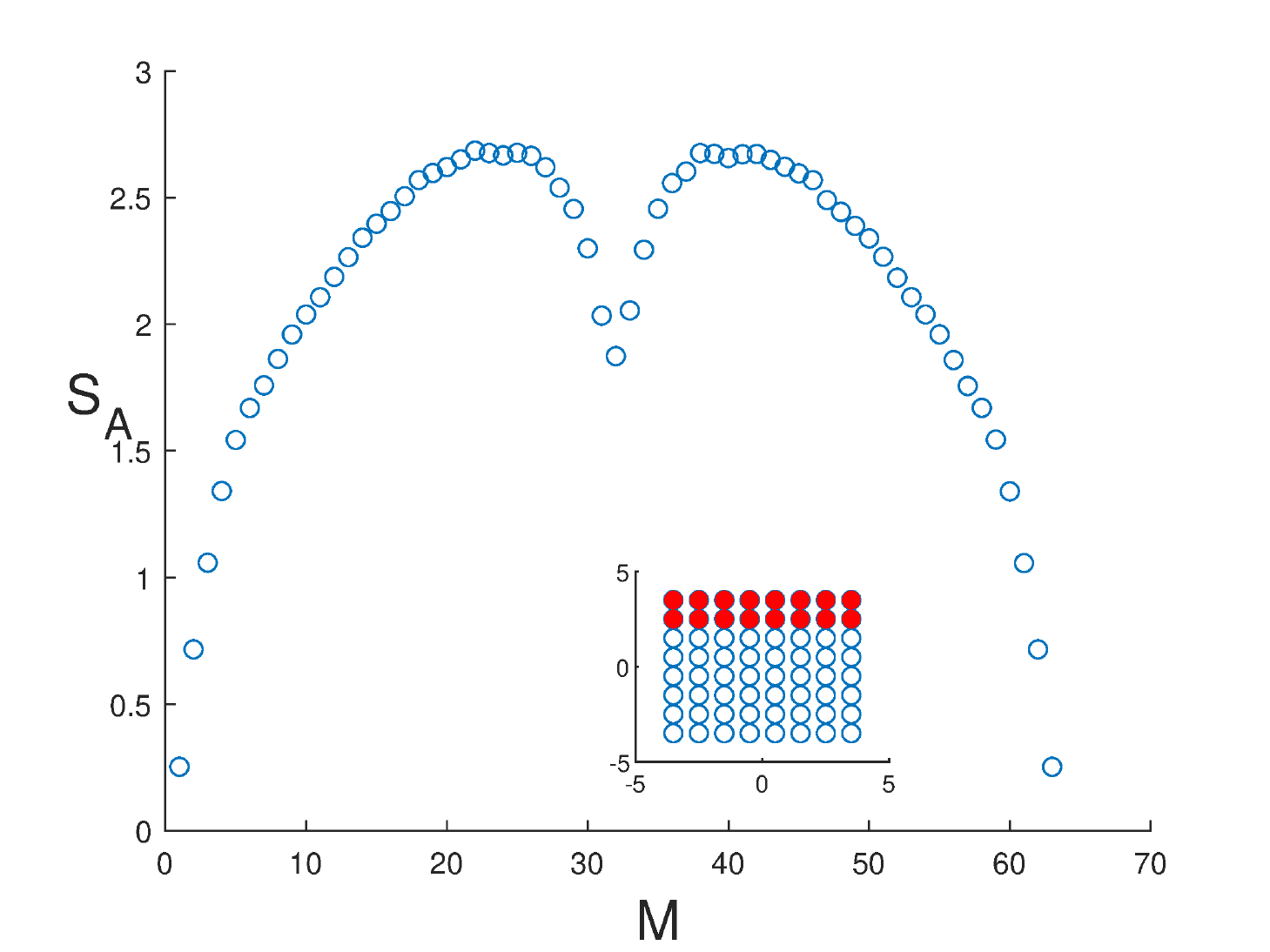}
    \hspace*{-51mm}\makebox[0mm]{\raisebox{32mm}{(a)}}\hspace*{51mm}
    \includegraphics[width=0.32\linewidth]{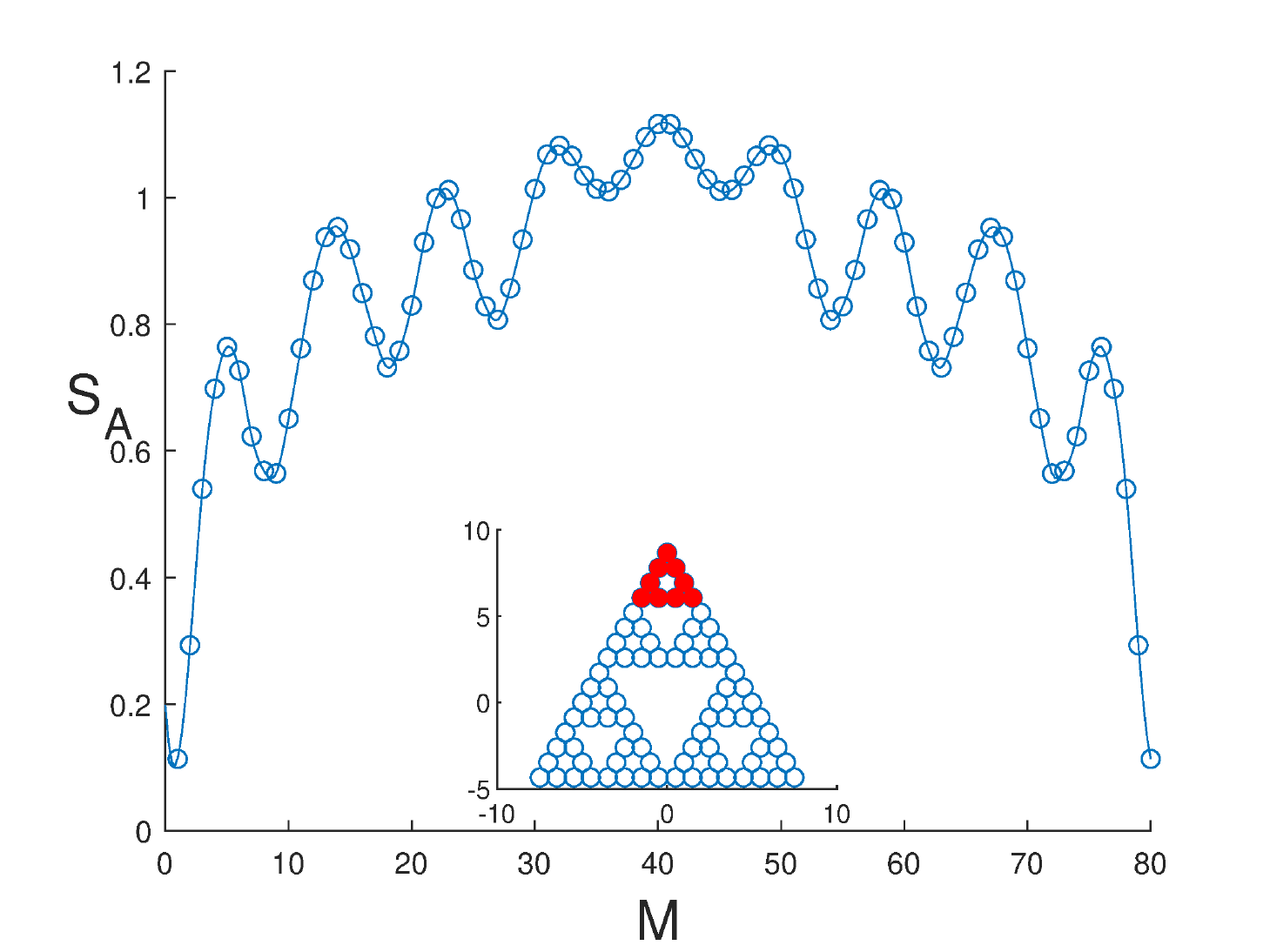}
    \hspace*{-51mm}\makebox[0mm]{\raisebox{32mm}{(b)}}\hspace*{51mm}
    \includegraphics[width=0.32\linewidth]{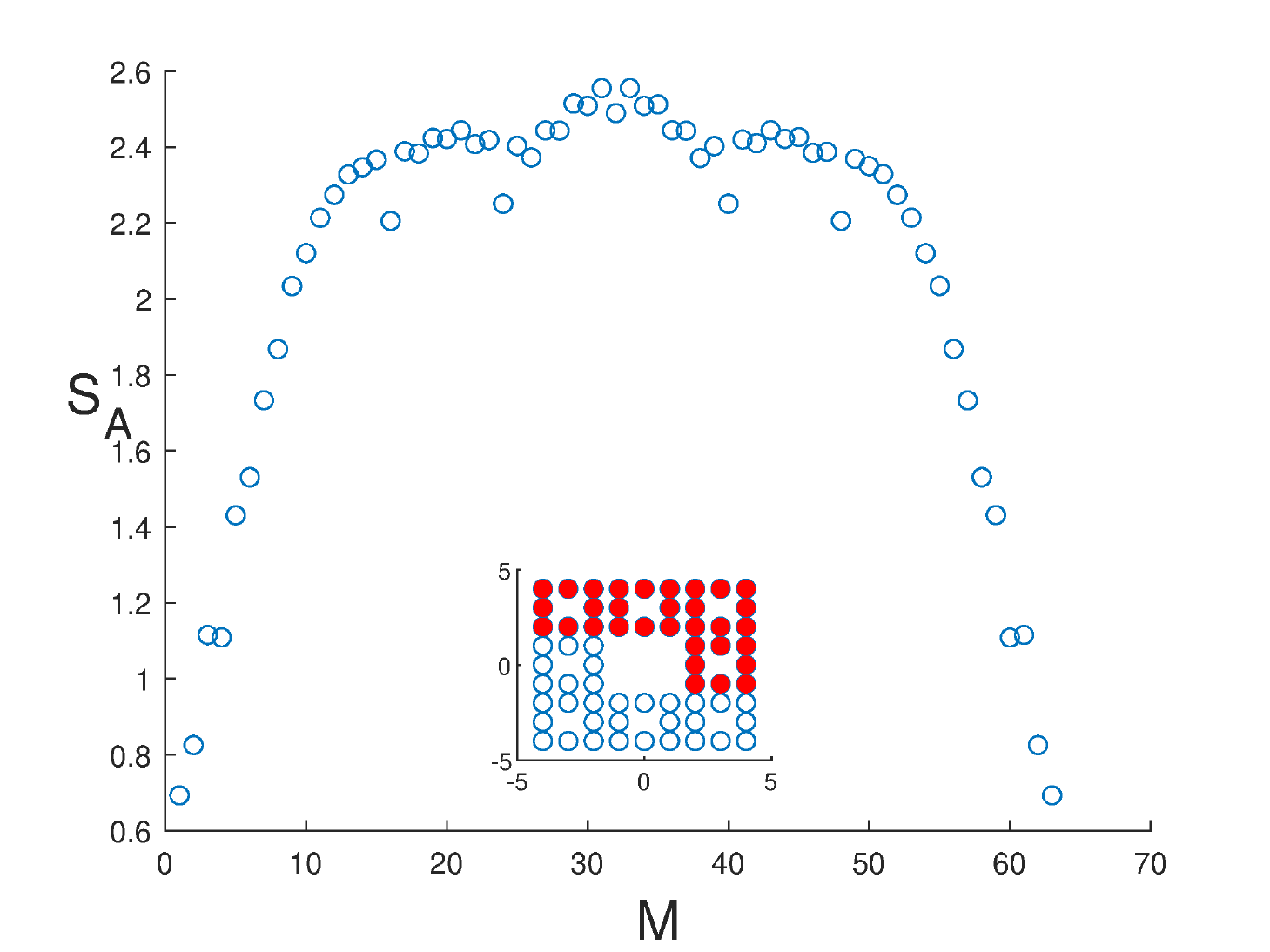}
    \hspace*{-51mm}\makebox[0mm]{\raisebox{32mm}{(c)}}\hspace*{51mm}\\
    \raisebox{40mm}{(d)}
    \includegraphics[width=0.40\linewidth]{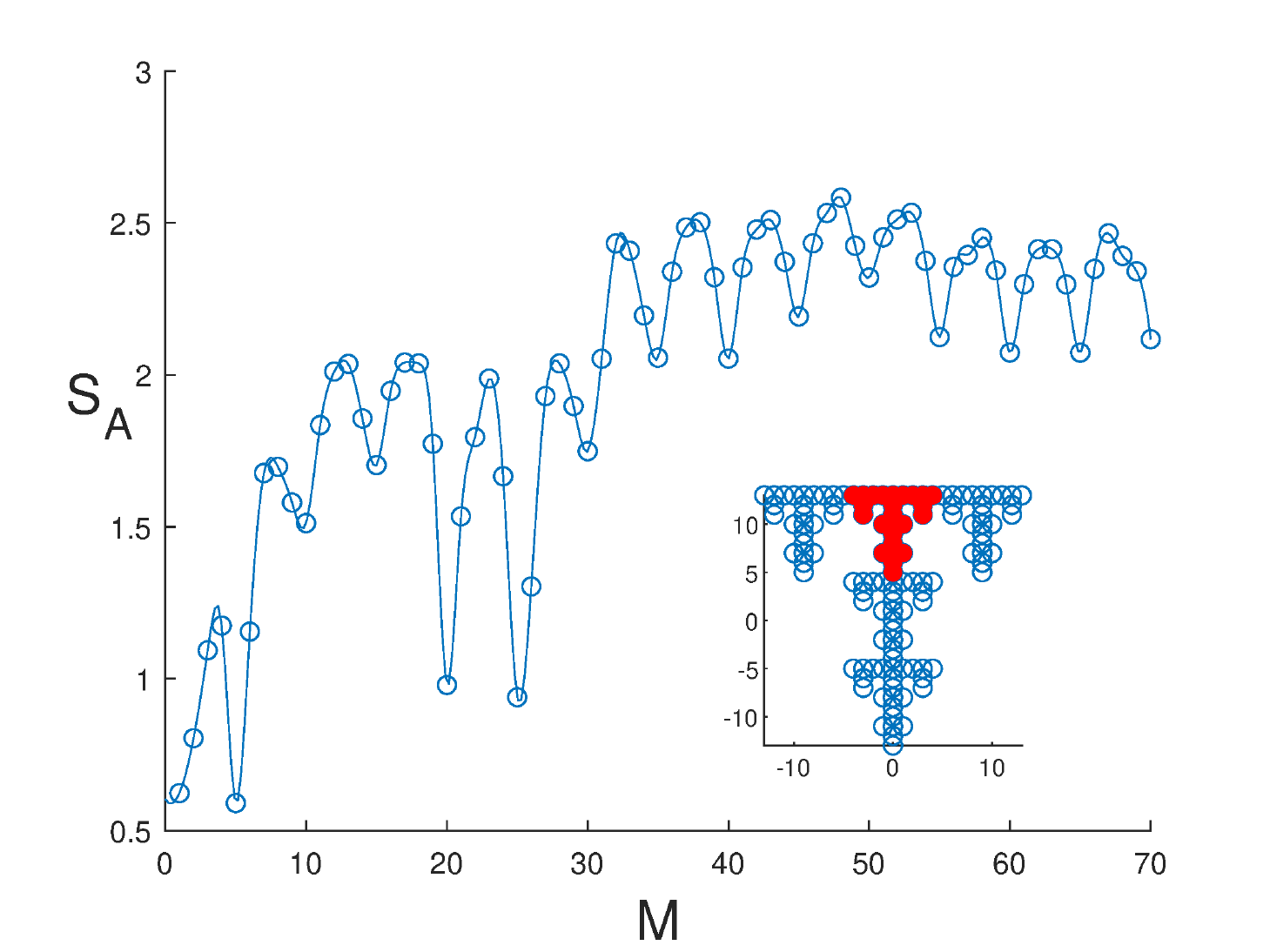}
    \raisebox{40mm}{(e)}
    \includegraphics[width=0.40\linewidth]{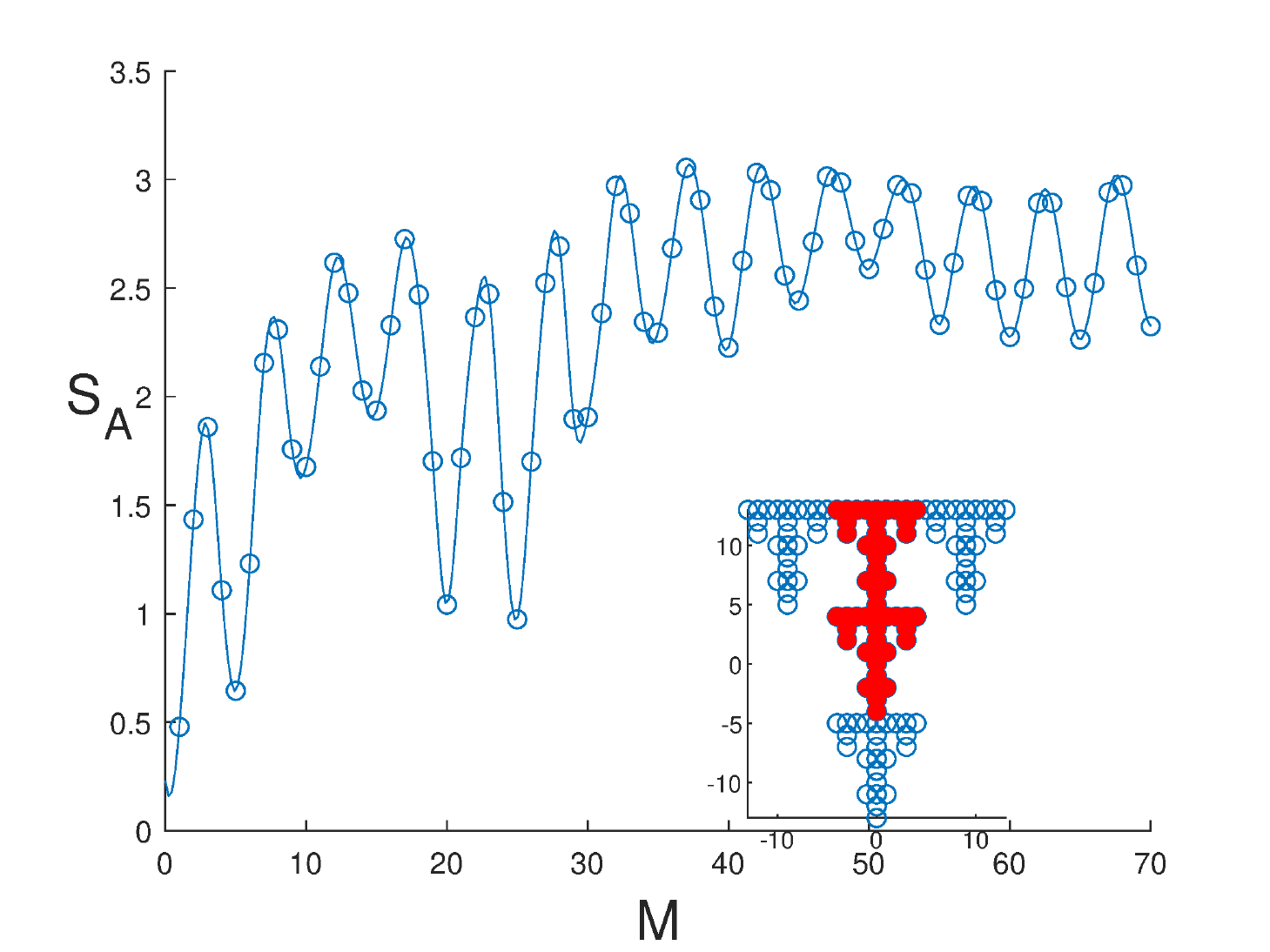}
    \caption{The entanglement entropy as a function of particle number $M$ for (a) an $8\times 8$ square lattice, (b) the Sierpinski triangle with $81$ sites, (c) the Sierpinski carpet with $64$ sites, and (d-e) the T-fractal with $125$ sites. For all the plots, $q=3$. Part $A$ is marked in red in the insets. The plots are symmetric around half-filling. The computations for the Sierpinski triangle were also done in \cite{LaughlinFractal}, and we include the results here for comparison. We observe that the entropy versus particle number shows periodic oscillations for the Sierpinski triangle and the T-fractal, but not for the square lattice and the Sierpinski carpet. For the T-fractal, the local minima lie at $5n$, where $n$ is a natural number.}
    \label{SvM}
\end{figure}

The entanglement entropy for an $8\times 8$ square lattice is shown in figure \ref{SvM}(a). A similar computation was done for the case of the cylinder in \cite{LaughlinLattice}. The entropy increases initially and then becomes constant and decreases. The second half of the curve is the mirror image of the first half. Initially, more particles means more entanglement and hence the increase makes sense. 

When the same state is hosted on a fractal lattice, however, the behaviour can be rather different. In \cite{LaughlinFractal}, the entropy as a function of $M$ was computed for the case of the Sierpinski triangle. We also recompute and plot the result for one specific case in figure \ref{SvM}(b). Instead of varying smoothly with $M$, the entropy shows oscillations with a precise period. The origin of these oscillations is, as of now, unclear.

The entropy plots for the T-fractal in figure \ref{SvM}(d-e) also show oscillations. In (d), part $A$ includes $25$ sites in the upper central part. The minima occur at $M=5n$, where $n$ is a natural number. This seems to follow the pattern observed in \cite{LaughlinFractal}, where the minima in the oscillations were found to be at $M=n N/N_{A}$ for the cases studied. In (e), however, we show the results for $N_{A}=50$. For this case, $N/N_{A}=5/2$ is not an integer. We nevertheless still find oscillations with the minima located at $M=5n$. This is also the case for the T-fractal with $25$ sites. Figure \ref{SvM}(c) shows results for the Sierpinski carpet. There are a few points that are lower than the others, but the oscillations, present in both the Sierpinski triangle and the T-fractal, are surprisingly absent for the Sierpinski carpet.

\section{Entropy scaling}\label{sec:scaling}

For systems with local interactions and a gapped energy spectrum, the entanglement entropy of the ground state is expected to follow an area law. If such a system has topological order, the intercept of the linear scaling with system size is a universal constant and is called the \textit{topological entanglement entropy} \cite{Kitaev2006,levin2006}. In \cite{LaughlinFractal}, it was found that the lattice Laughlin state on a fractal does \textit{not} obey an area law. In fact, for a specific bipartition, an approximately logarithmic scaling was found. Here, we investigate the scaling further for different bipartitions in order to gain deeper insights into the entanglement of such a strongly correlated state on the Sierpinski triangle. The scaling is done by increasing the fractal generation and the number of particles, while keeping $M/N$, $N_A/N$, and the number of nearest neighbour bonds crossing the border of the bipartition constant. With this scaling, the entropy should be independent of system size if the system follows the area law. We instead find that the entropy generally increases with system size. We consider only the Sierpinski triangle here, because the number of sites increases faster with the generation for the Sierpinski carpet and the T-fractal, which limits the number of data points we can obtain.  

Our results are shown in figure \ref{scaling}. The chosen bipartitions are illustrated in (a-c) and (g-i). When the generation is increased by one, each lattice site turns into three lattice sites, which are assigned to the same bipartition as the original site. In this process the number of nearest neighbor bonds crossing the border of the bipartition remains constant, e.g.\ $4$ for the bipartition in (a). In other words, the area remains constant. We also increase the number of particles such that the average particle density $M/N$ remains constant. The behavior of the entropy with system size as well as linear fits are shown in (d-f) and (j-l). While the entropy increases with system size for all cases, showing that the area law in not fulfilled, the scaling behavior depends on the choice of bipartition. In some cases, the data points follow a scaling that is close to logarithmic, while in other cases the increase of the entropy with subsystem size is faster, and in some cases the entropy is closer to scaling with the volume of the subsystem, at least for the system sizes considered. The results hence reveal a quite complicated behavior of the entropy. 

\def\mylabel#1{\raisebox{33mm}{#1}\hspace{-1mm}}
\begin{figure*}
     \mylabel{(a)}
      \includegraphics[width=0.29\linewidth]{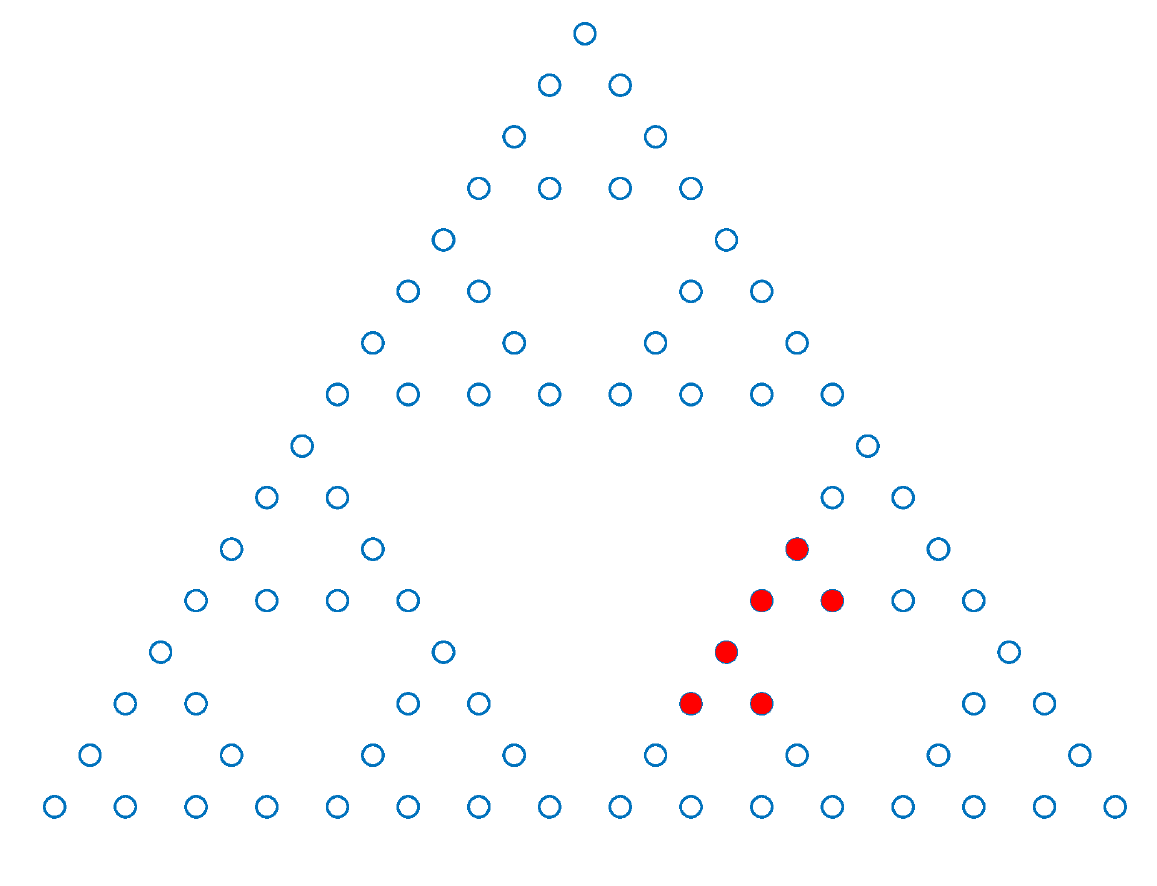}
    \mylabel{(b)} 
    \includegraphics[width=0.29\linewidth]{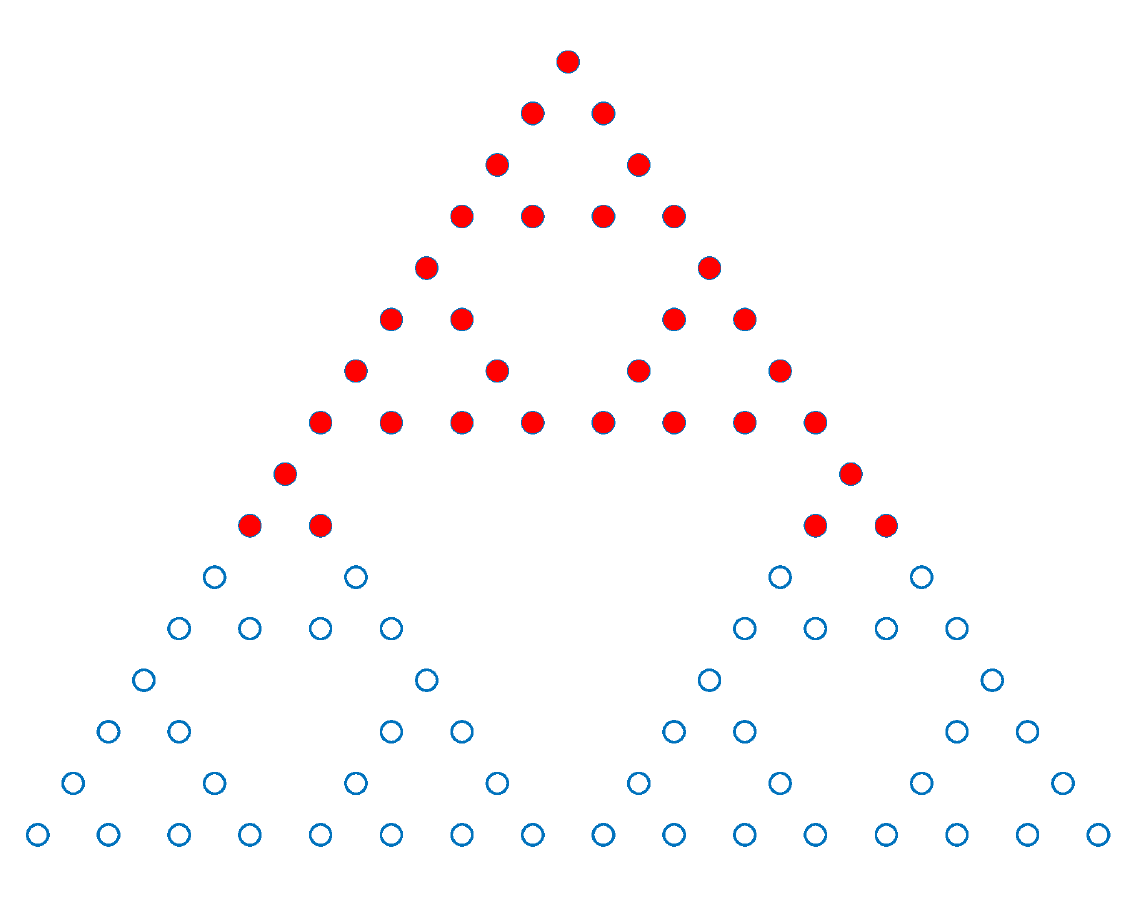}
   \mylabel{(c)}
    \includegraphics[width=0.29\linewidth]{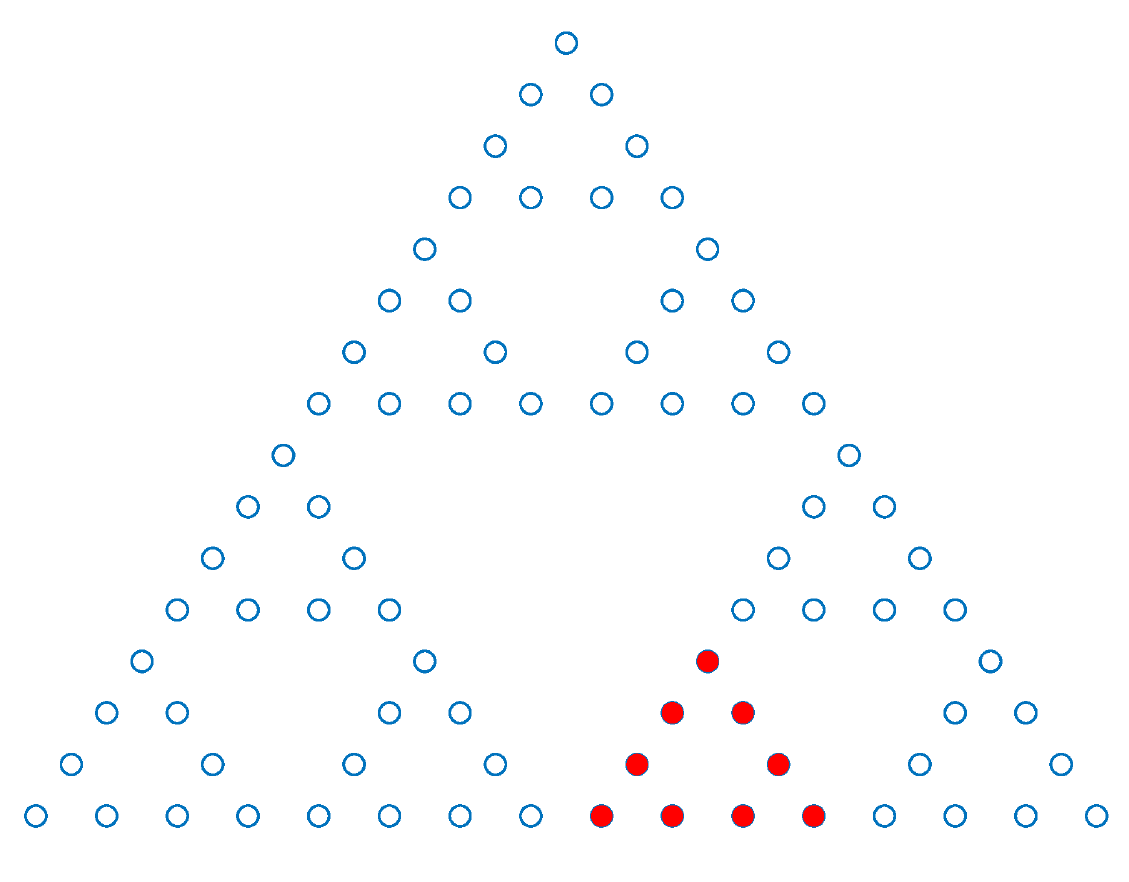}
    \mylabel{(d)}
    \includegraphics[width=0.29\linewidth]{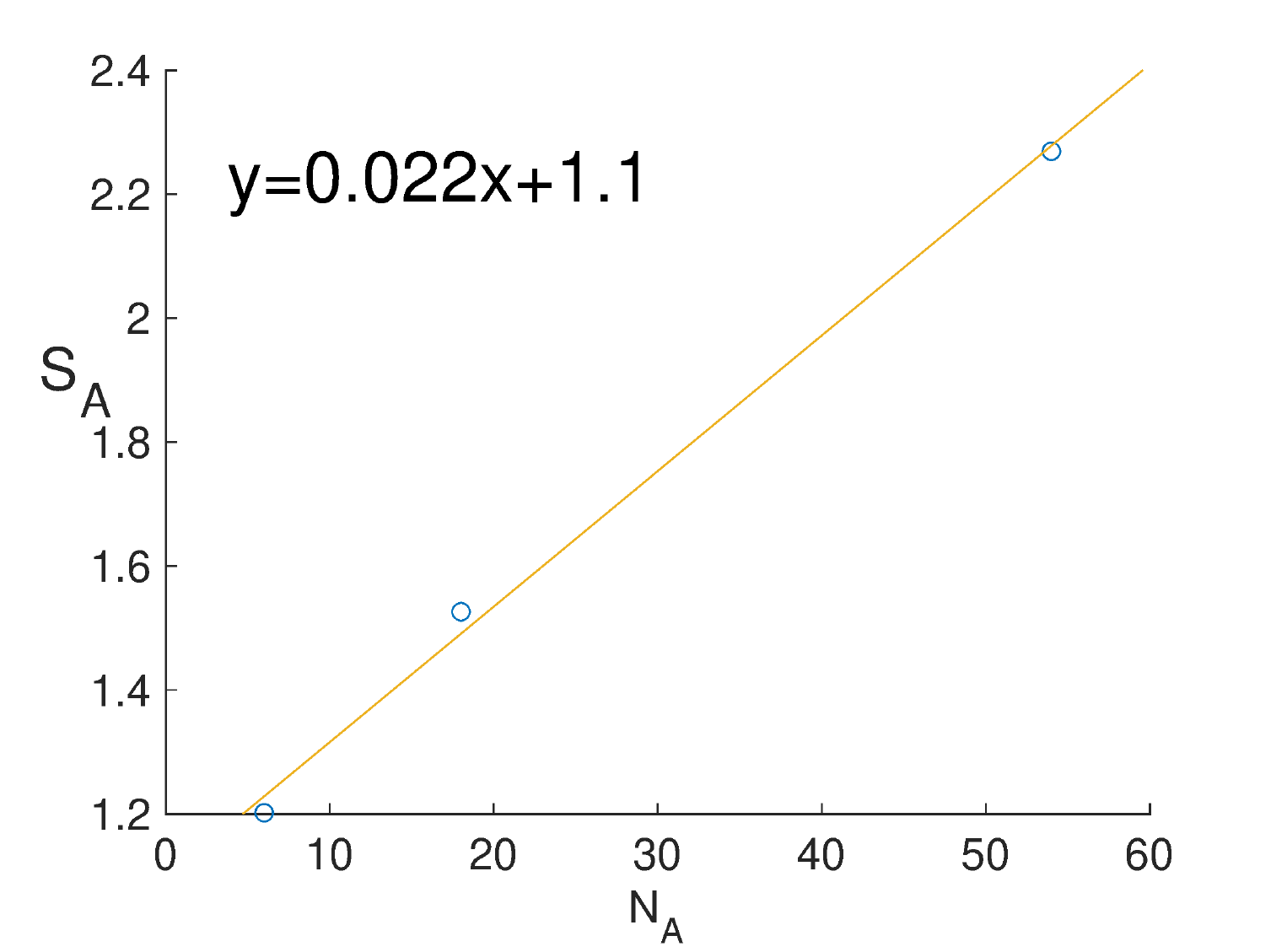}
    \mylabel{(e)}
     \includegraphics[width=0.29\linewidth]{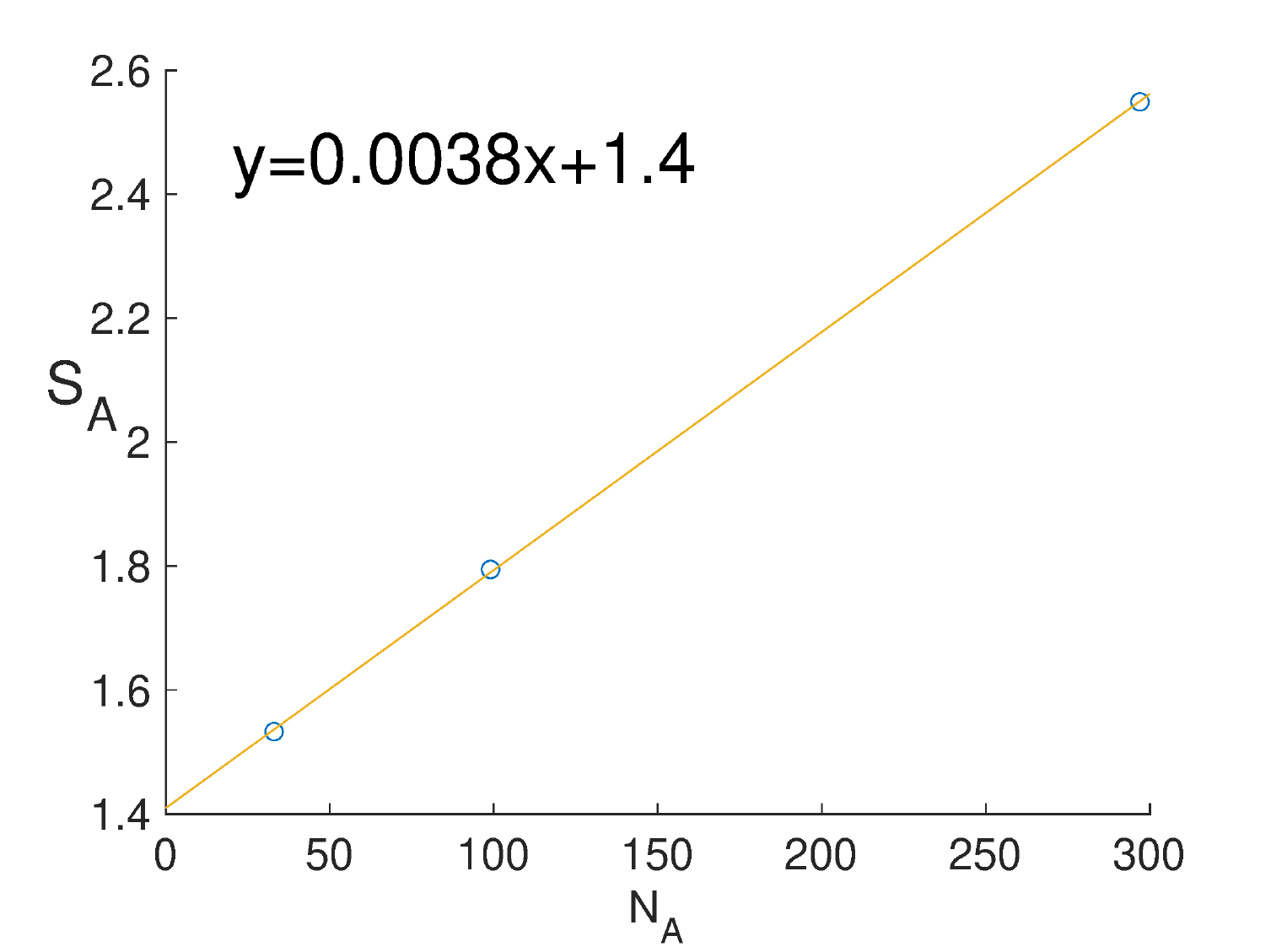}
     \mylabel{(f)}
      \includegraphics[width=0.29\linewidth]{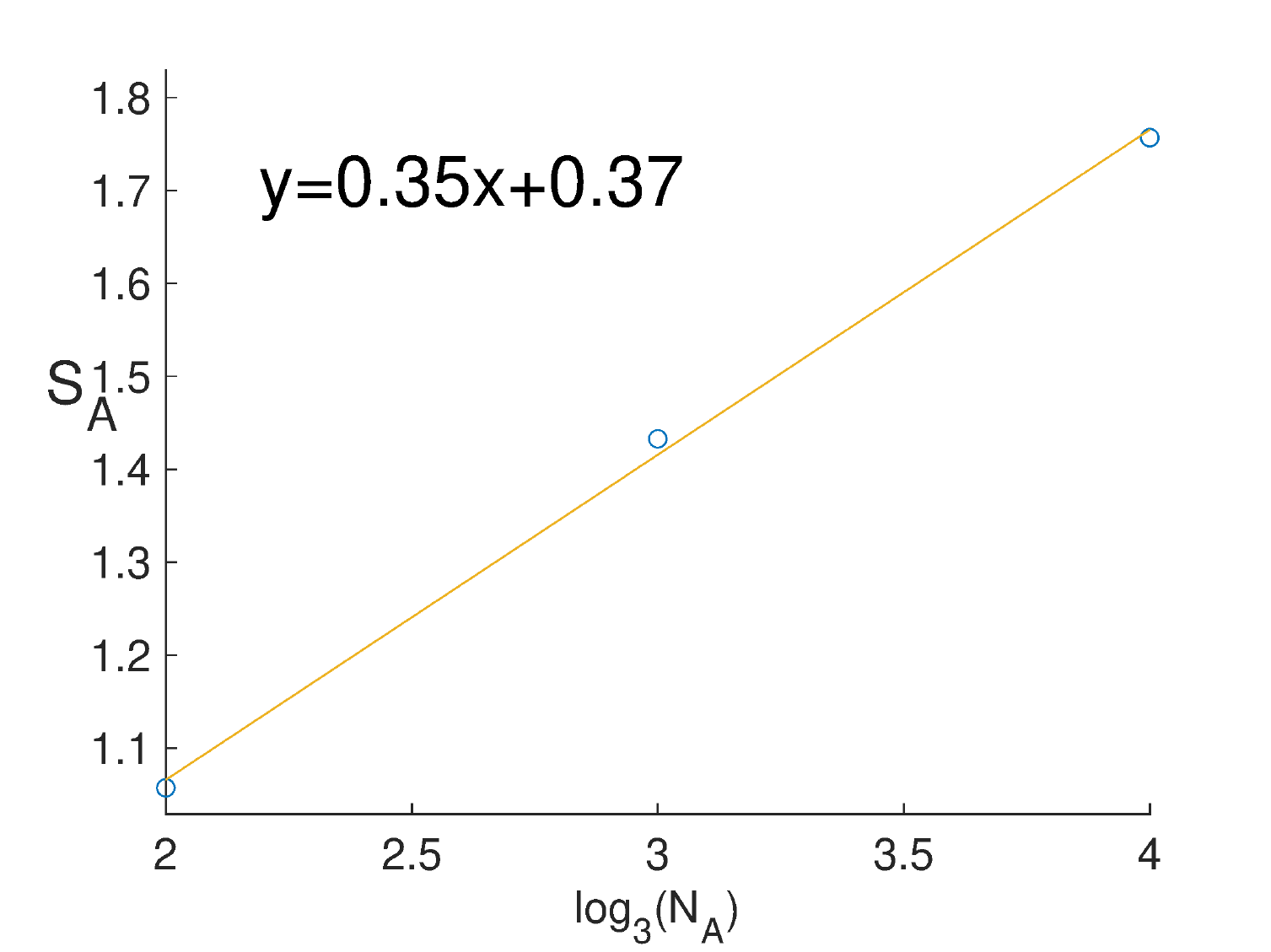}
   \mylabel{(g)}
    \includegraphics[width=0.29\linewidth]{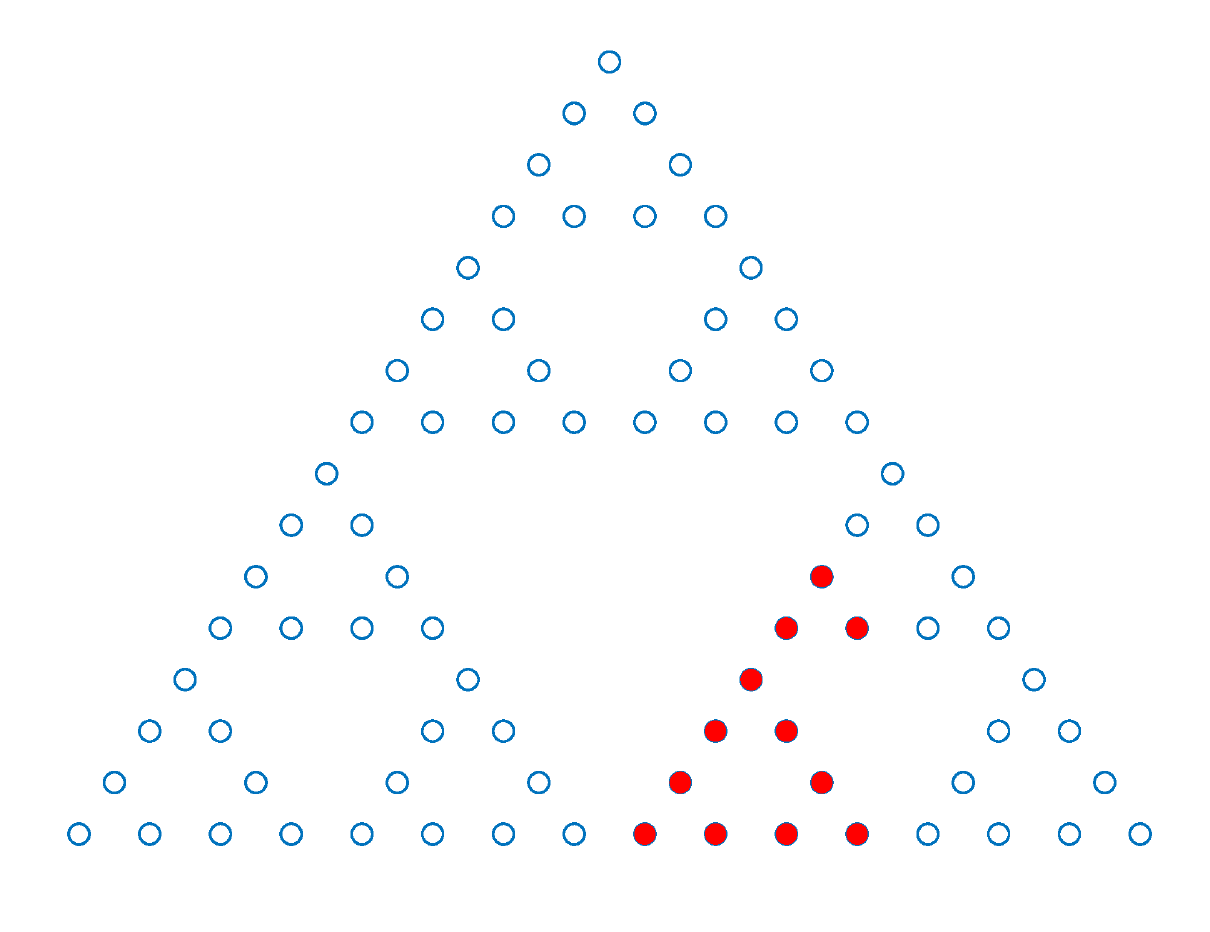}
    \mylabel{(h)}
   \includegraphics[width=0.29\linewidth]{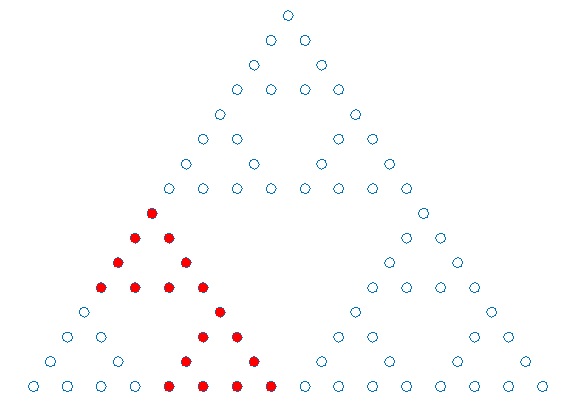}
    \mylabel{(i)} 
     \includegraphics[width=0.29\linewidth]{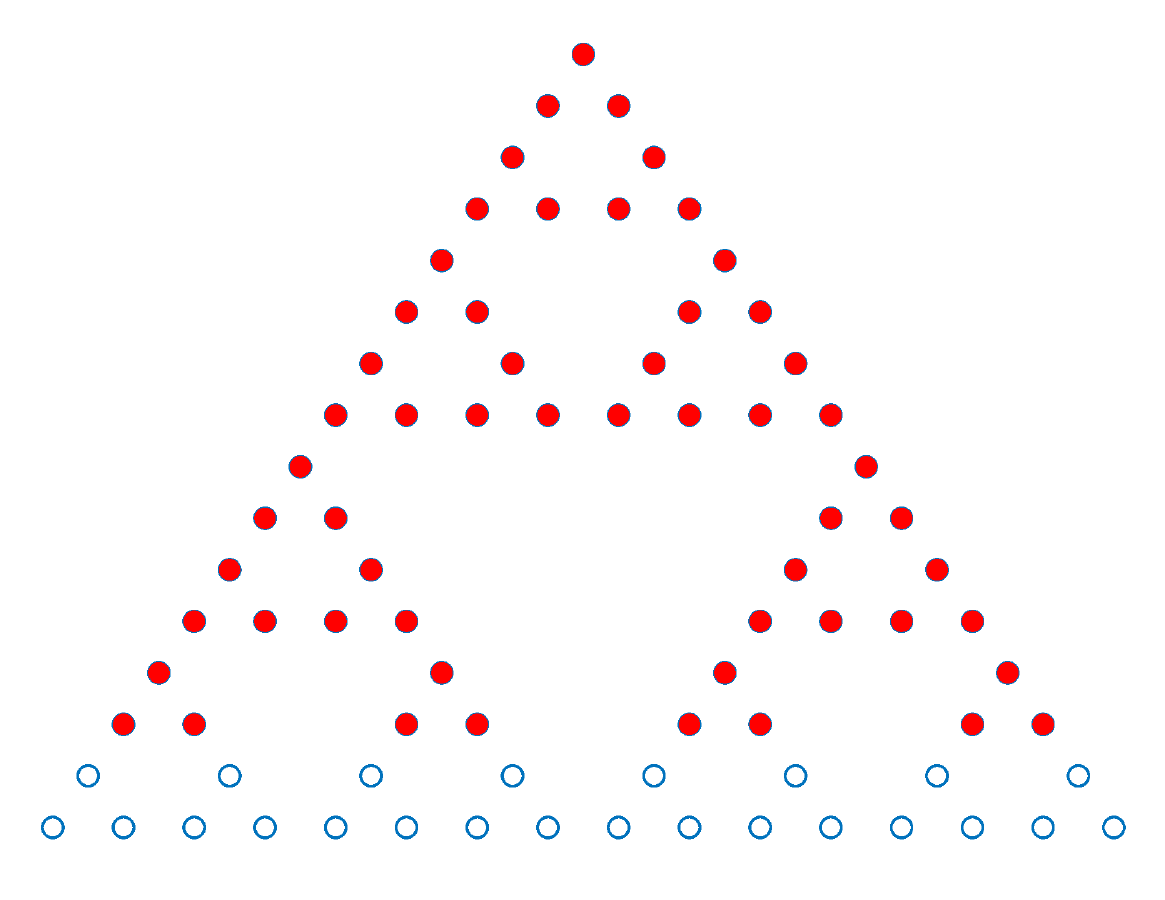}
    \mylabel{(j)}
   \includegraphics[width=0.29\linewidth]{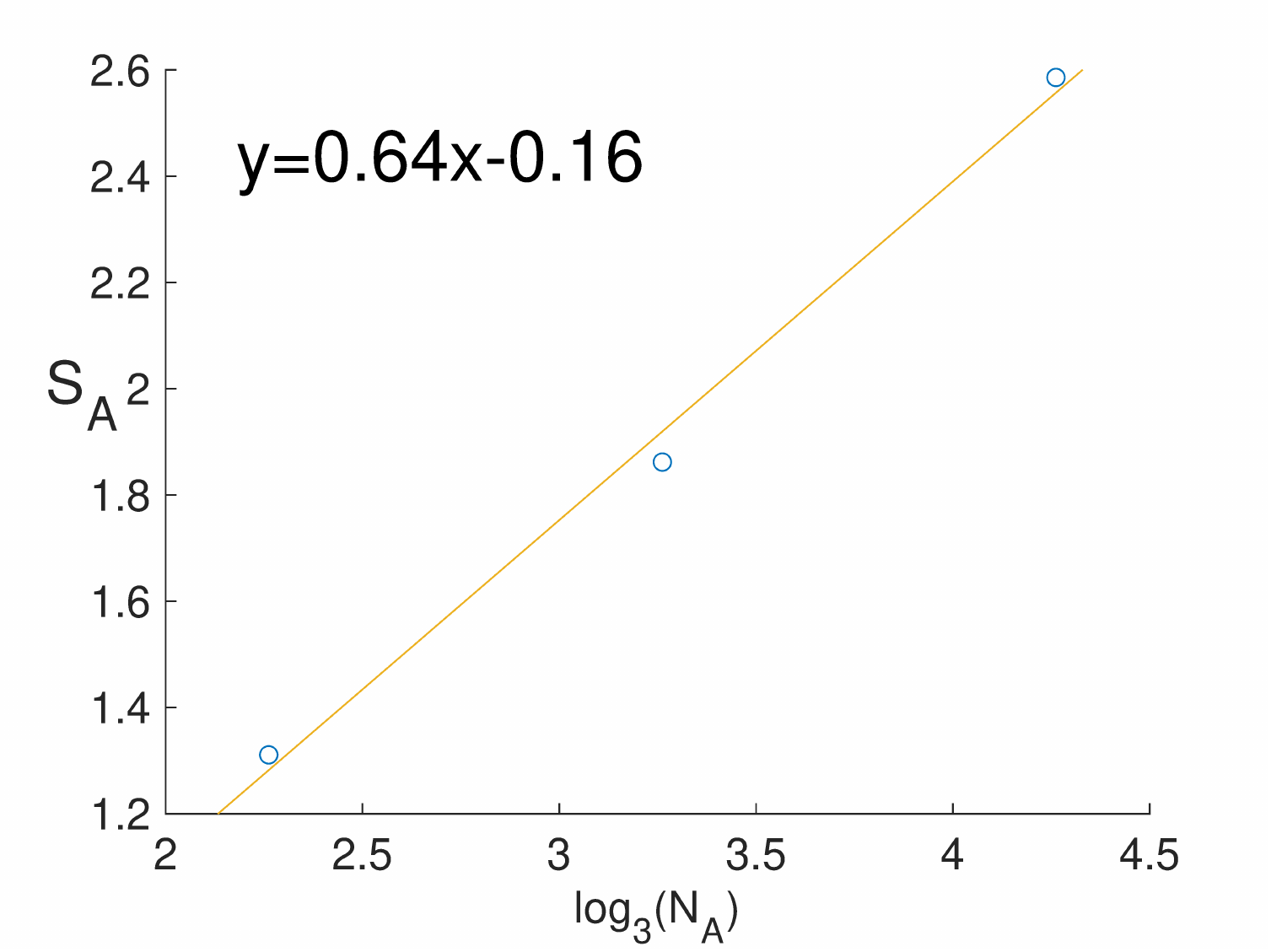}
    \mylabel{(k)}
     \includegraphics[width=0.29\linewidth]{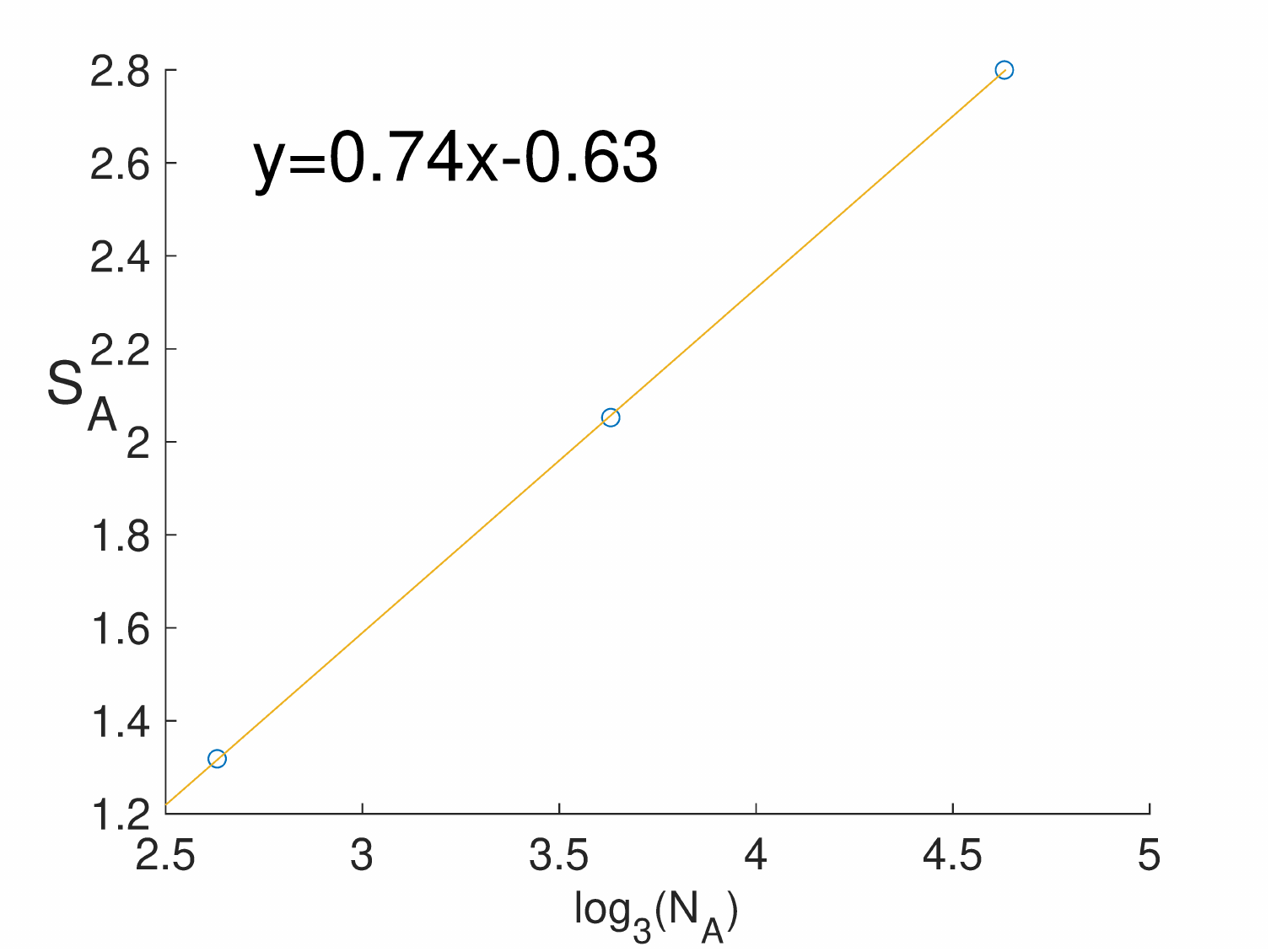}
     \mylabel{(l)}
    \includegraphics[width=0.29\linewidth]{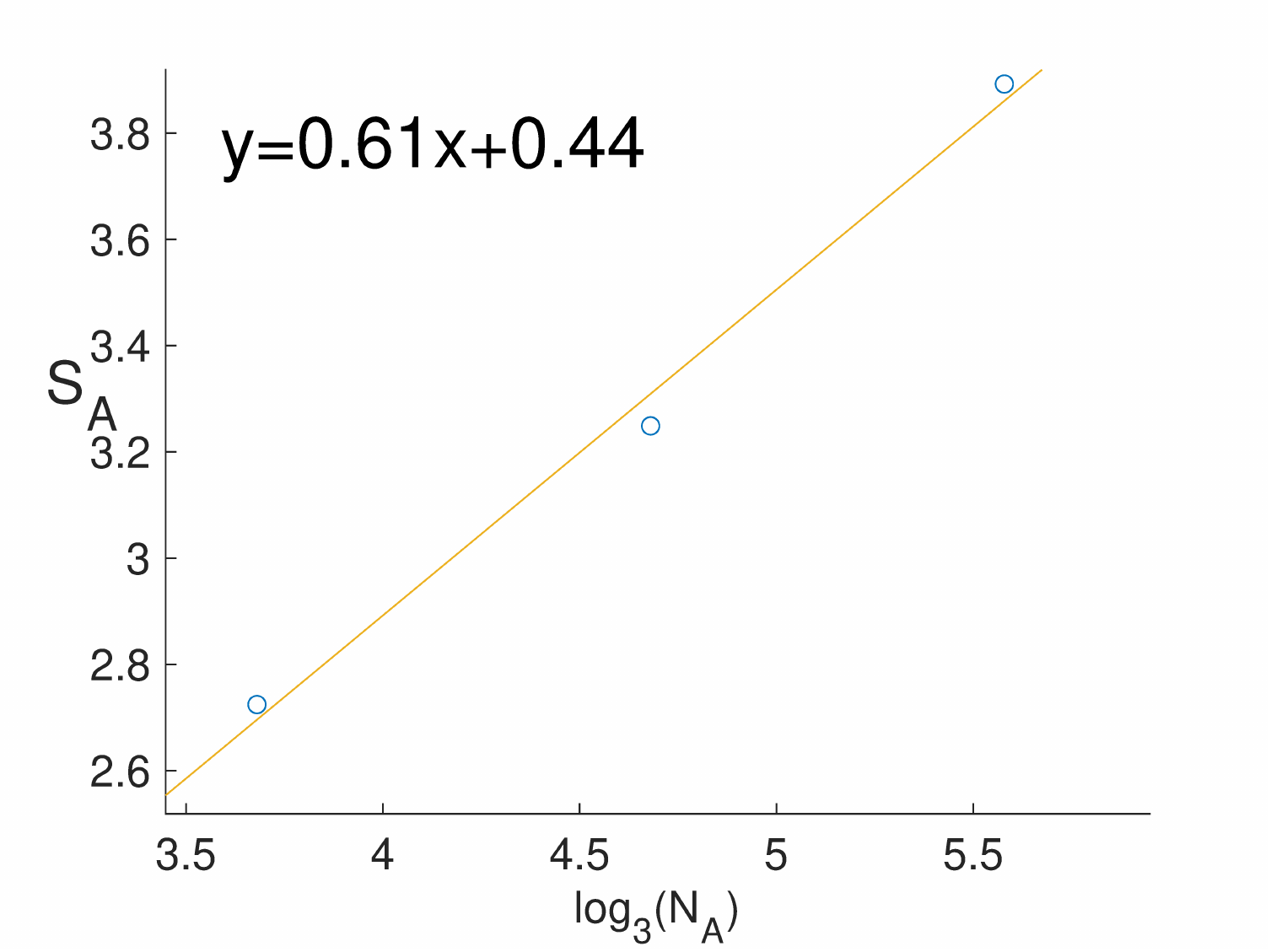}    
   \caption{Scaling of the Renyi entropy with subsytem size for various bipartitions on a Sierpinski triangle. For each bipartition, we increase the system size by increasing the generation. In doing so, we keep $N_A/N$ and the shape of the subsystem unchanged. The average density of particles $M/N$ is also unchanged. The figure above each fit shows the chosen subsystem (marked red) for the fractal with $N=81$ sites, and the other two data points are for $N=243$ and $N=729$. The lines are linear fits with the slope and intersection given in the plots.}
   \label{scaling}
\end{figure*}

\section{Conclusion} \label{sec:conclusion}

Motivated by our findings in \cite{LaughlinFractal}, we studied the properties of the lattice Laughlin state on different kinds of fractal lattices with different Hausdorff dimensions and different ramification numbers, and we found a variety of non-trivial and interesting features.

First, we studied the particle densities on the fractal lattice as a function of the particle number $M$. We observed that for most values of $M$, the densities at different sites vary with position, but at half-filling, the particle density is one-half for all sites due to particle-hole symmetry. For the Sierpinski triangle, we also observed that the variation of the particle density with position is slightly reduced at $\eta=1$. For most values of the particle number, the densities of the Sierpinski triangle and Sierpinski carpet fall in two distinct groups corresponding to the sites in the corners and sites in the `bulk'. The densities at the `bulk' sites form a thick central band around the mean value, while the corner sites deviate from this band. Thus, even though a fractal typically does not offer a distinct concept of `bulk' and `edge', one sees a clear distinction in the density patterns, but only for large enough system sizes. 

We inserted both quasiholes and quasiparticles into the Laughlin state on fractal lattices and studied how their presence affect the particle density. We found that the quasiparticles are about the same size as the quasiholes. We also showed how the size of the anyons varies with position in the fractal lattices. For the considered fractal lattices, we observed that the size of the anyons tends to increase when the Hausdorff dimension decreases. 

An important difference between periodic lattices in two dimensions and the Sierpinski triangle and carpet is that the Sierpinski triangle and carpet have inner edges in addition to the outer edge. We constructed trial states describing edge states on both the outer and the inner edges of the Sierpinski triangle and carpet.

It has been found previously \cite{LaughlinFractal} that the entropy shows oscillations as a function of the particle number for the Sierpinski triangle, but not for the square lattice. Here, we showed that oscillations are also present for the T-fractal, but not for the Sierpinski carpet. This suggests that the ramification number could play a role for the oscillations, as the ramification number is infinite for the square lattice and the Sierpinski carpet and finite for the Sierpinski triangle and the T-fractal. 

We also investigated how the entanglement entropy scales with subsystem size for the Sierpinski triangle. We took a sequence of subsystems in such a way that the generation of the fractal increased without changing the number of nearest neighbour bonds crossing the border of the bipartition. We observed for several different bipartitions that the entropy increases with subsystem size, and in some cases the increase is faster than logarithmic. This is starkly different from the two-dimensional case, where the area law applies. 

Our results show that topologically ordered systems, when hosted on fractal lattices, can show a variety of non-trivial properties which have no counterpart in periodic lattices in two dimensions.

\section*{Acknowledgments}
This work has been supported by the Independent Research Fund Denmark under grant number 8049-00074B.

\appendix

\section{Particle-hole transformation of the lattice Laughlin state} \label{sec:appendix}

In this section, we show explicitly how the lattice Laughlin state (\ref{laughlin}) transforms when particles and holes are exchanged, i.e., $n_{i} \to 1 - n_{i}$, $\forall i \in \{1,2,\ldots,N\}$. When doing this operation, we should note that $\eta=q\sum_i n_i/N$, so $\eta \to q - \eta$.

We first note that the delta function 
\begin{equation}
\delta_{\sum_i(1-n_i),M}=\delta_{N-\sum_i n_i,M}=\delta_{\sum_i n_i,N-M}\equiv\delta_{n,N-M}
\end{equation}
picks out terms with $N-M$ particles. Next we consider the factors 
\begin{equation}
\tilde{\psi}(n_1,\ldots,n_N)\equiv\prod_{i<j} (z_i-z_j)^{q n_i n_j}
\prod_{k\neq l} (z_k-z_l)^{-\eta n_l}. 
\end{equation}
They have the property
\begin{eqnarray}
\fl\tilde{\psi}(1-n_1,\ldots,1-n_N)=
\prod_{i<j} (z_i-z_j)^{q (1-n_i)(1-n_j)}
\prod_{k\neq l} (z_k-z_l)^{-(q-\eta)(1 - n_l)}\\
=\prod_{i<j} (z_i-z_j)^{(q -qn_i-qn_j+qn_{i}n_{j})}
\prod_{k\neq l} (z_k-z_l)^{(-q+\eta+qn_{l} - \eta n_l)}\\
= \tilde{\psi}(n_{1},\ldots,n_{N}) \prod_{i<j} (z_i-z_j)^{(q-qn_i-qn_j)}
\prod_{k\neq l} (z_k-z_l)^{(-q+\eta+qn_{l})}\\
= \tilde{\psi}(n_{1},\ldots,n_{N}) \prod_{i<j} (z_i-z_j)^{(q-qn_i-qn_j)}\nonumber\\
\phantom{=}\times \prod_{k<l} (z_k-z_l)^{(-q+\eta+qn_{l})} \prod_{k>l} (z_k-z_l)^{(-q+\eta+qn_{l})}. 
\end{eqnarray}
We relabel the indices $k \to i$  and $ l \to j $
\begin{eqnarray}
 = \tilde{\psi}(n_{1},\ldots,n_{N}) \prod_{i<j} (z_i-z_j)^{(q-qn_i-qn_j)}\nonumber\\
\phantom{=}\times \prod_{i<j} (z_i-z_j)^{(-q+\eta+qn_{j})} \prod_{i>j} (z_i-z_j)^{(-q+\eta+qn_{j})}\\ 
= \tilde{\psi}(n_{1},\ldots,n_{N}) \prod_{i<j} (z_i-z_j)^{(-qn_i+\eta)} \prod_{i>j} (z_i-z_j)^{(-q+\eta+qn_{j})}. 
\end{eqnarray}
Now we switch $i$ and $j$ in the second product only
\begin{eqnarray}
= \tilde{\psi}(n_{1},\ldots,n_{N}) \prod_{i<j} (z_i-z_j)^{(-qn_i+\eta)} \prod_{i<j} (z_j-z_i)^{(-q+\eta+qn_{i})}\\
 = \tilde{\psi}(n_{1},\ldots,n_{N}) \prod_{i<j} (z_i-z_j)^{(-qn_i+\eta)}\nonumber\\
\phantom{=}\times\prod_{i<j} (-1)^{(-q+\eta+qn_{i})}\prod_{i<j}(z_i-z_j)^{(-q+\eta+qn_{i})}.\\
=\tilde{\psi}(n_{1},\ldots,n_{N}) \prod_{i<j} (z_i-z_j)^{(2\eta-q)}
\prod_{i<j} (-1)^{(-q+\eta)}
\prod_{k} (-1)^{qn_{k}(N-k)}.
\label{transform}
\end{eqnarray}
Note that the factor $\prod_{i<j}(-1)^{(-q+\eta)}\prod_{i<j}(z_{i}-z_{j})^{(2\eta-q)}$ is independent of $n_{i}$. Note also that for the special case of $M=N/2$, the Kronecker delta remains the same, and the wavefunction only changes by the sign factors in (\ref{transform}) as $q=2\eta$.

Finally, we consider the normalization factor. Considering a system with $M$ particles, we must have
\begin{equation}
\frac{1}{|C_M|^2}=\sum_{n_1,\ldots,n_N} \delta_{n,M} |\tilde{\psi}(n_1,\ldots,n_N)|^2
\end{equation}
By changing summation variables and using (\ref{transform}) we get
\begin{eqnarray}
\fl\sum_{n_1,\ldots,n_N} \delta_{n,M} |\tilde{\psi}(n_1,\ldots,n_N)|^2
=\sum_{n_1,\ldots,n_N} \delta_{n,N-M} |\tilde{\psi}(1-n_1,\ldots,1-n_N)|^2\nonumber\\
=\sum_{n_1,\ldots,n_N} \delta_{n,N-M} |\tilde{\psi}(n_1,\ldots,n_N)\prod_{i<j} (z_i-z_j)^{(2\eta-q)}
\prod_{i<j} (-1)^{(-q+\eta+qn_{i})}|^2\nonumber\\
=\sum_{n_1,\ldots,n_N} \delta_{n,N-M} |\tilde{\psi}(n_1,\ldots,n_N)\prod_{i<j} (z_i-z_j)^{(2\eta-q)}
\prod_{i<j} (-1)^{(-q+\eta)}|^2.
\end{eqnarray}
Considering a system with $N-M$ particles, however, we must also have
\begin{equation}
\frac{1}{|C_{N-M}|^2}=\sum_{n_1,\ldots,n_N} \delta_{n,N-M} |\tilde{\psi}(n_1,\ldots,n_N)|^2.
\end{equation}
Combining the last three equations, we conclude
\begin{equation}\label{C_M}
|C_{N-M}|=|C_{M}|\,\left|\prod_{i<j} (z_i-z_j)^{(2\eta-q)}
\prod_{i<j} (-1)^{(-q+\eta)}\right|.
\end{equation}

\subsection{Density}
We will here denote the expectation value of the number operator in the state with $M$ particles by $\langle \hat{n}_i\rangle_M$. Consider
\begin{eqnarray}\label{denstrans}
\fl 1-\langle \hat{n}_i\rangle_M
=|C_M|^2\sum_{n_1,\ldots,n_N} (1-n_i) \delta_{n,M}|\tilde{\psi}(n_1,\ldots,n_N)|^2\nonumber\\
=|C_M|^2\sum_{n_1,\ldots,n_N} n_i \delta_{n,N-M}|\tilde{\psi}(1-n_1,\ldots,1-n_N)|^2\nonumber\\
=|C_{N-M}|^2\sum_{n_1,\ldots,n_N} n_i \delta_{n,N-M}|\tilde{\psi}(n_1,\ldots,n_N)|^2
=\langle \hat{n}_i\rangle_{N-M}.
\end{eqnarray}
At the second equality sign, we changed summation variables, and at the third equality sign, we used (\ref{transform}) and (\ref{C_M}). Subtracting $(N-M)/N$ on both sides of this equation gives
\begin{equation}
\langle \hat{n}_i\rangle_{N-M}-\frac{N-M}{N}=-\left(\langle \hat{n}_i\rangle_M-\frac{M}{N}\right)
\end{equation}
This expresses the symmetry observed in the plots in section \ref{sec:density}. For the special case of $M=N-M=N/2$, (\ref{denstrans}) yields $\langle \hat{n}_i\rangle_{N/2}=1/2$ for all $i$.

\subsection{Entanglement entropy}

In this section, we show analytically that the Renyi entanglement entropy for the system with $M$ particles is the same as the Renyi entanglement entropy for the system with $N-M$ particles. We shall here denote the Renyi entanglement entropy for the system with $M$ particles by $S_{A,M}$.  

We first observe that
\begin{equation}
\frac{\tilde{\psi}(1-n,1-m')\tilde{\psi}(1-n',1-m)}{\tilde{\psi}(1-n,1-m)\tilde{\psi}(1-n',1-m')}
=\frac{\tilde{\psi}(n,m')\tilde{\psi}(n',m)}{\tilde{\psi}(n,m)\tilde{\psi}(n',m')},
\end{equation}
because all the factors $(-1)^{qn_{k}(N-k)}$ in the numerator with and without primes cancel corresponding factors in the denominator. We can hence write
\begin{eqnarray}
\fl \rme^{-S_{A,M}}=
|C_M|^4\sum_{n,m,n',m'}\delta_{(n,m),M}
\delta_{(n',m'),M}\delta_{(n',m),M}\delta_{(n,m'),M}\frac{\tilde{\psi}(n,m')\tilde{\psi}(n',m)}{\tilde{\psi}(n,m)\tilde{\psi}(n',m')}\nonumber\\
\phantom{=}\times|\tilde{\psi}(n,m)|^2 |\tilde{\psi}(n',m')|^2\nonumber\\
=|C_M|^4\sum_{n,m,n',m'}\delta_{(n,m),N-M}\delta_{(n',m'),N-M}\delta_{(n',m),N-M}\delta_{(n,m'),N-M}\nonumber\\
\phantom{=}\times\frac{\tilde{\psi}(n,m')\tilde{\psi}(n',m)}{\tilde{\psi}(n,m)\tilde{\psi}(n',m')}|\tilde{\psi}(1-n,1-m)|^2 |\tilde{\psi}(1-n',1-m')|^2\nonumber\\
=|C_{N-M}|^4\sum_{n,m,n',m'}\delta_{(n,m),N-M}\delta_{(n',m'),N-M}\delta_{(n',m),N-M}\delta_{(n,m'),N-M}\nonumber\\
\phantom{=}\times\frac{\tilde{\psi}(n,m')\tilde{\psi}(n',m)}{\tilde{\psi}(n,m)\tilde{\psi}(n',m')}|\tilde{\psi}(n,m)|^2 |\tilde{\psi}(n',m')|^2
=\rme^{-S_{A,N-M}}.
\end{eqnarray}
From this we conclude that $S_{A,M}=S_{A,N-M}$.

\section*{References}
\providecommand{\newblock}{}

\end{document}